\documentclass[preprint,review,1p,times,12pt]{elsarticle}
\usepackage{setspace}
\makeatletter
\def\ps@pprintTitle{%
 \let\@oddhead\@empty
 \let\@evenhead\@empty
 \def\@oddfoot{\centerline{\thepage}}%
 \let\@evenfoot\@oddfoot}
\makeatother
\usepackage[utf8]{inputenc}
\usepackage{amsmath,nicefrac}
\usepackage{mathtools}
\usepackage{wasysym}
\usepackage{arydshln}
\RequirePackage{doi}
\usepackage{hyperref}
\hypersetup{
    colorlinks=true,
    linkcolor=blue,
    pdfborderstyle={/S/U}}
\usepackage[scr=rsfso]{mathalfa}
\usepackage{graphicx}
\usepackage{lineno}
\let\oldalign\align
\let\oldendalign\endalign
\renewenvironment{align}
  {\linenomathNonumbers\oldalign}
  {\oldendalign\endlinenomath}
\begin{document}

\begin{frontmatter}

\title{Exact closed-form and asymptotic expressions for the electrostatic force between two conducting spheres}
\author{Shubho Banerjee}

\author{Thomas Peters}

\author{Nolan Brown}

\author{Yi Song}

\address{Department of Physics, Rhodes College, Memphis, TN 38112}

\begin{abstract}
We present exact closed-form expressions and complete asymptotic expansions for the electrostatic force between two charged conducting spheres of arbitrary sizes. Using asymptotic expansions of the force we confirm that even like-charged spheres attract each other at sufficiently small separation unless their voltages/charges are the same as they would be at contact. We show that for sufficiently large size asymmetries, the repulsion between two spheres $\textit{increases}$ when they separate from contact if their voltages or their charges are held constant. Additionally, we show that in the constant voltage case, this like-voltage repulsion can be further increased and  maximised though an optimal $\textit{lowering}$ of the voltage on the larger sphere at an optimal sphere separation.
\end{abstract}



\end{frontmatter}
\section{Introduction}\label{sec:introduction}
The interaction of two charged conducting spheres is a fundamental problem in electrostatics that has a history of more than one hundred years. Lord Kelvin~\cite{Thomson1872-jh}, Poisson,  Kirchoff, Maxwell~\cite{Maxwell1954-ka}, and Russell~\cite{Russell1927-lr} are among those who have worked on this problem. Due to its nontrivial nature, the problem continues to generate interest to this date~\cite{Lekner2011-pz,Lekner2012-pr,Lekner2012-ua,Kolikov2012-zq,Meyer2015-fd,Banerjee2017-xl,Lekner2016-df,Banerjee2019-vf}. The simple and ubiquitous geometry of spheres makes the analytical results relevant in a wide variety of applications where electrostatic forces play an important role such as the interaction of raindrops in clouds~\cite{Harrison2020-xo}, dust and powder interactions~\cite{Feng2003-wu,Cordero2017-dk}, cellular and molecular interactions~\cite{Varadwaj2017-rb}, atomic force microscopy~\cite{Hudlet1998-fe,Law2002-go}, etc.

The calculations and analysis of the force becomes difficult as the two spheres approach each other due to charge polarisation on each sphere. In the classical solution using the method of images for example, an ever increasing number of images need to be included for convergence and the solutions fail completely when the two spheres touch each other. In the near limit asymptotic solutions provide an alternative series to calculate the force accurately with a short computation time.
However, in this asymptotic limit, the force expansions are known only to the first order in the sphere surface-to-surface separation~\cite{Lekner2012-pr}. Due to the difficulty of this analysis, only recently has it been shown by Lekner~\cite{Lekner2012-ua} that two like-charged spheres attract each other at sufficiently small separation unless they have the exact same charge ratio as they would at contact. 

In this paper we present a complete asymptotic analysis of the force in the near limit to {\it all} orders in sphere separation. Additionally, we derive exact closed-form expressions for the force using the $q$-digamma function~\cite{Krattenthaler1996-ev}.
Our analysis confirms the attraction between like-charged  spheres and provides new results on how the electrostatic force varies with size asymmetry. In particular, we find that at large size asymmetries  the behaviour of the force is radically different from the behaviour at small asymmetries.

\begin{figure}
\centering
\scalebox{0.204}{\includegraphics{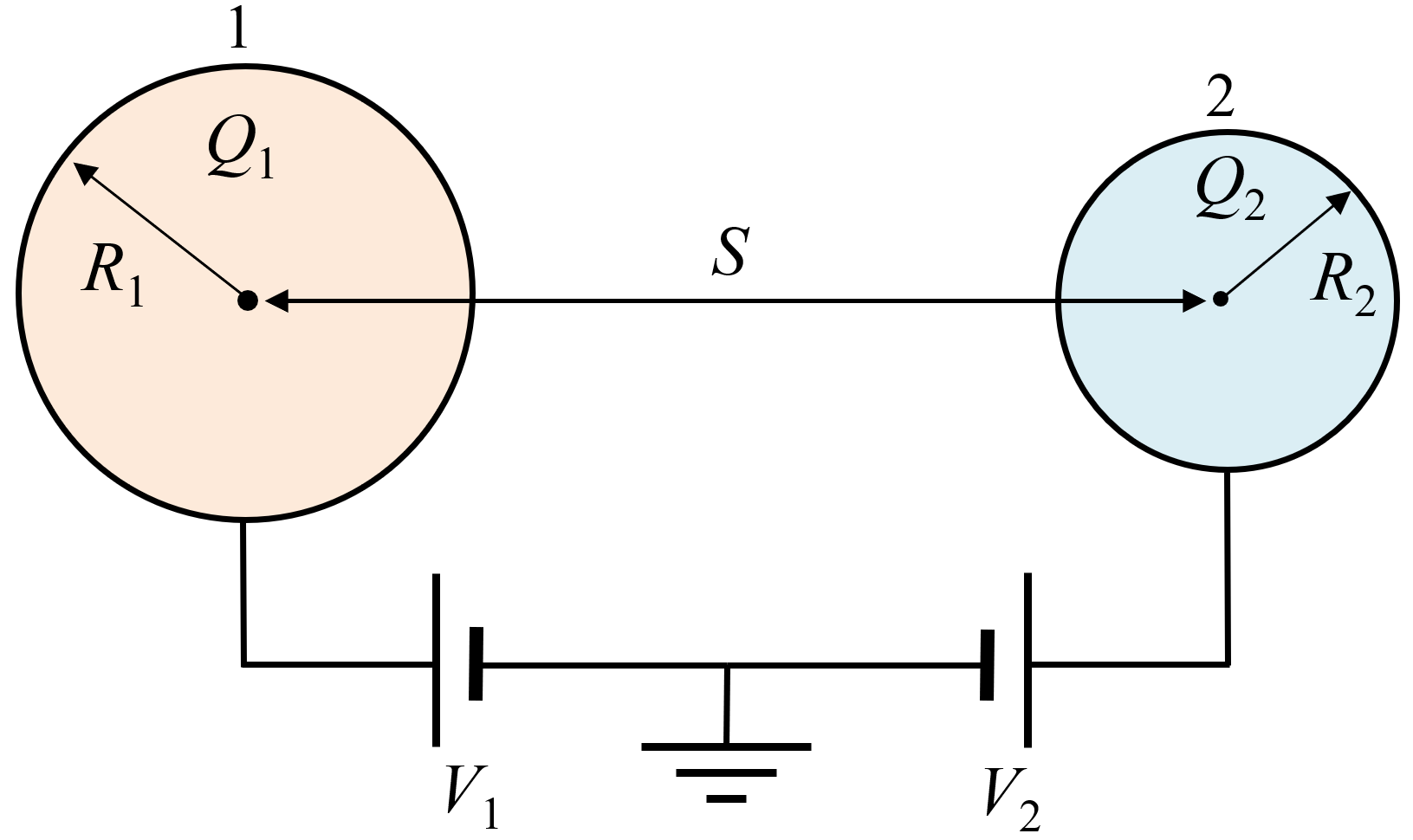}}
\caption{Two conducting spheres $1$  and $2$ are held at voltages $V_1$ and $V_2$ respectively. The permittivity of the surrounding medium is $\epsilon$. The spheres acquire charges $Q_1$  and $Q_2$ to maintain their given voltages in the presence of each other. These charges vary with separation. If the batteries are disconnected at some given charge, then the voltages of the two spheres vary with separation.}
\label{fig:TwoSpheres}
\end{figure}

The setup of the problem is shown in Fig.~\ref{fig:TwoSpheres}. Two spheres 1 and 2 with radii $R_1$ and $R_2$ are at a distance $S$ between their centres and are given the voltages $V_1$ and $V_2$ respectively. The charges $Q_1$ and $Q_2$ acquired by spheres are related to their given voltages through the matrix equation
\begin{align} \label{eq:Q=CV}
	\begin{pmatrix}
		Q_1 \\
		Q_2
	\end{pmatrix} \equiv
	\begin{pmatrix}
		C_{11}   & C_{12} \\ 
		C_{21}   & C_{22}
	\end{pmatrix}
	\begin{pmatrix}
		V_1 \\
		V_2
	\end{pmatrix} ,
\end{align}
where $C_{11}$ and $C_{22}$ are the self capacitances of the two spheres and $C_{12}=C_{21}$ is their mutual capacitance. These capacitance coefficients are geometric properties of the interaction and depend only on $R_1$, $R_2$, and $S$~\cite{Maxwell1954-ka,Smythe1968-rd}. 

In a recent paper~\cite{Banerjee2019-vf} we presented exact closed-form and asymptotic solutions to all orders in sphere separation for these capacitance coefficients. In Sec.~\ref{sec:capacitance} we present a summary of these results along with the well known classical solutions since the knowledge of the capacitance coefficients is vital to the analysis presented in this paper.

The basic formalism for calculating the force between any two conductors (see Art. 93 in Ref.~\cite{Maxwell1954-ka} for example), starts by writing the overall electrostatic energy of the system as
\begin{align}
W_V
=&\frac{1}{2} \left[C_{11} V_1^2 + 2 C_{12} V_1 V_2+C_{22} V_2^2\right]
\end{align}
when the voltages of the two spheres are held constant. The mutual electrostatic force between the two spheres is related to the variation of this energy with distance
\begin{align} \label{eq:fvdefinition}
F_V=\frac{dW_V}{dS}=\frac{1}{2} \left[\frac{dC_{11}}{dS} V_1^2 + 2 \frac{dC_{12}}{dS} V_1 V_2+\frac{dC_{22}}{dS} V_2^2\right].
\end{align}
In Sec.~\ref{sec:capacitance_derivatives} we carry out the derivatives of all three forms of the capacitance coefficients discussed in Sec.~\ref{sec:capacitance}. These derivatives are the essential components required for calculation of the force.

In Sec.~\ref{sec:voltage} we calculate and plot this force at constant voltage for some typical sphere size ratios. For spheres with sufficiently large size asymmetries, we show that repulsion at contact initially {\it increases} with distance. Additionally,  we show that the maximum repulsion between two spheres happens not when both spheres have the same voltage but when voltage of the larger sphere is optimally {\it lower} and the spheres are at an optimal non-zero separation. Our analysis reveals numerical values for the size asymmetry thresholds above which these anomalous behaviours of the force are observed.

If the spheres have a constant amount of charge instead, then inverting the charge voltage relation gives
\begin{align}
\begin{pmatrix} V_1\\ V_2 
\end{pmatrix}
\,=\,\begin{pmatrix} 
C_{11} & C_{12} \\
C_{12} & C_{22} 
\end{pmatrix}^{-1}
\begin{pmatrix} 
Q_1\\
Q_2
\end{pmatrix}\,\equiv\,
\begin{pmatrix} 
P_{11} & P_{12} \\
P_{12} & P_{22} 
\end{pmatrix}
\begin{pmatrix} 
Q_1\\
Q_2
\end{pmatrix},
\end{align}
where $P_{ij}$ are the coefficients of potential~\cite{Maxwell1954-ka}. The electrostatic energy of the system and the electrostatic force are given by (see Art. 93 in Ref.~\cite{Maxwell1954-ka})
\begin{align}
W_Q
=&\frac{1}{2} \left[P_{11} Q_1^2 + 2 P_{12} Q_1 Q_2+P_{22} Q_2^2\right],~~~~ F_Q =
  -\frac{dW_Q}{dS}.
\end{align}
By analysing the coefficients $P_{ij}$, Lekner showed that two charged spheres always attract each other unless they have the exact same charge ratio as they would at contact. In this paper we confirm this result by the explicit calculation of the asymptotic force to all orders in sphere separation. In addition, we show that above a certain size asymmetry, similar to the constant voltage case, the repulsion between two charged spheres {\it increases}  at first when they separate  from contact even as their charges remain fixed.

\section{Capacitance coefficients} \label{sec:capacitance}
As discussed in Sec.~\ref{sec:introduction}, the key to understanding the force between the two spheres lies in examining the geometrical dependence of the capacitance coefficients. In this section we present a recap of the classical solutions~\cite{Maxwell1954-ka,Smythe1968-rd} and some recently published asymptotic and closed-form capacitance results~\cite{Banerjee2019-vf} that are required for accurate force calculations.

\subsection{Preliminary definitions}

In writing down expressions for the capacitance coefficients, it is convenient to write the following two geometric definitions:
\begin{equation}
s \equiv \frac{S}{{R_1}+{R_2}}
\end{equation}
is the dimensionless distance between the spheres. When the two spheres are about to touch, $s\rightarrow 1$. The dimensionless surface separation of the two spheres is $s-1$ in units of $R_1+R_2$.
In addition to $s$ we define a second parameter that measures the asymmetry in the radii of the two spheres
\begin{equation}
r \equiv \frac{{R_1}-{R_2}}{{R_1}+{R_2}}.
\end{equation}
For equal sized spheres, $r=0$, and  when the second sphere is much smaller in size, $r\rightarrow 1$. Interchanging the two spheres takes $r \rightarrow -r$.

The quantities $s$ and $r$ are a complete description of the geometry of the problem. In terms of $s$ and $r$ it is convenient to define
\begin{align} 
\label{eq:alpha}\alpha&= \left( \frac{\sqrt{s^2-r^2}- \sqrt{s^2-1}}{\sqrt{1-r^2}}\right)^{2},
\end{align}
which scales the entire distance range between the two spheres to the unit interval. When the two spheres are about to touch $\alpha \rightarrow 1$, and when the spheres move very far away $\alpha \rightarrow 0$. 

The relation between distance variables $s$ and $\alpha$ in Eq.~(\ref{eq:alpha}) can be inverted to give
\begin{equation}\label{eq:s-of-alpha}
s=\frac{\sqrt{(1+\alpha)^2 -r^2(1-\alpha)^2}}{2 \sqrt{\alpha }}.
\end{equation}
To write the capacitances in compact form it is useful to define
\begin{align}
\lambda = \frac{(1-r^2)}{2s}\left(\frac{1-\alpha^2}{\alpha}\right)=\frac{2(1-r^2)\sinh \mu}{\sqrt{1 - r^2 \tanh^2 \mu}}
\end{align}
and
\begin{equation}
 \frac{1}{2}-x =y-\frac{1}{2}=\frac{\tanh^{-1}(r\tanh \log\!\sqrt{\alpha}\hspace{0.1em})}{\log \alpha}=\frac{\tanh^{-1}(r\tanh \mu )}{2\mu}, 
\end{equation}
where $\mu = - \log \sqrt \alpha$ or we may write $\alpha = \exp(-2\mu)$. Similar to $s-1$, $\mu \rightarrow 0$ when the spheres are about to touch. Note that the closely related variable $u = -\log {\alpha}$ is used by Lekner to express the capacitance coefficients using the Abel-Plana formula~\cite{Lekner2011-pz}. Maxwell first defined this $u$ variable (see $\varpi$ in Chapter XI of Ref.~\cite{Maxwell1954-ka}) and Russell~\cite{Russell1927-lr} uses this variable in his papers as well.

The size asymmetry $r$, along with one distance variable $\alpha$, $\mu$, or $s$, completely specifies the geometry of the problem. For our numerical calculations we use $r$ and $\alpha$ as our main quantities, and express $\mu$, $s$, $\lambda$, $x$ and $y$ in terms of these two quantities. We choose $\alpha$ as the main distance variable in our numerical calculations because it is easily converted to $s$ or $\mu$ through the simple relations discussed above; this choice optimises the  computation time for the force plots presented in this paper.

To gain more insight into the interdependence of these quantities it is useful to look at their behaviour when the spheres are about to touch each other.  For small $\mu$ we have
\begin{align}
s-1&= \frac{ (1 - r^2)}{2}\mu^2  + \frac{(1 + 2 r^2 - 3 r^4)}{24} \mu^4  + {\cal O}(\mu^6),\\
\lambda &= 2(1- r^2)\,\mu  +\frac{(1+2 r^2-3 r^4)}{3}\mu ^3  + {\cal O}(\mu^5),
\end{align}
and 
\begin{align}
\frac{1}{2}-x&=y-\frac{1}{2}=\frac{r}{2}-\frac{r (1-r^2)}{6} \mu ^2  +\frac{ r\, (2-5 r^2+3 r^4)}{30} \mu ^4+ {\cal O}(\mu^6).
\end{align}
For spheres of equal size, both $x$ and $y$ are equal to $1/2$ independent of the sphere separation. The variable $\lambda$ is the derivative of $2s$ with respect to $\mu$ (see Sec.~\ref{sec:capacitance_derivatives}).

\subsection{Capacitance expressions}
It is useful to define the dimensionless capacitance coefficients for the two spheres as $c_{ij} \equiv C_{ij}/2\pi \epsilon(R_1+R_2)$.
With the definitions above, the classical solutions using the method of images can be written in the following infinite Lambert series~\cite{Banerjee2017-wk} form 
\begin{align} \label{eq:c11-Classical}
c_{11} &=\lambda\,\sum_{n=0}^\infty \frac{\alpha^{n+x}}{1-\alpha^{2n+2x}},~~~~~~
c_{22} =\lambda\,\sum_{n=0}^\infty \frac{\alpha^{n+y}}{1-\alpha^{2n+2y}},
\end{align}
and
\begin{align}
\label{eq:c12-Classical}
c_{12}
&=-\lambda \sum_{n=1}^\infty \frac{\alpha^n}{1-\alpha^{2n}}.
\end{align}
The series above (with minor differences) were first given by Kirchhoff~\cite{Maxwell1954-ka} and discussed in Ref.~\cite{Banerjee2019-vf} in their current form. We include terms through $n=10$ and limit the range to $\alpha \le 0.15$ with errors of order $10^{-10}$ or less for the calculations presented in this paper. Note that these series are defined for all $\alpha \neq 1$.

Using the sum of the Lambert series in terms of the $q$-digamma function~\cite{Krattenthaler1996-ev,Banerjee2017-wk} the capacitance coefficients can be written in closed-form  as~\cite{Banerjee2019-vf}
\begin{align}\label{eq:c11closedform}
c_{11}  &= \frac{\lambda}{4\mu}\, \left[{\psi_{\alpha^2}(x)-2\psi_{\alpha}(x)+\log \tfrac{1+\alpha}{1-\alpha}}\right],\\
\label{eq:c22closedform} c_{22}  &= \frac{\lambda}{4\mu}\, \left[{\psi_{\alpha^2}(y)-2\psi_{\alpha}(y)+\log \tfrac{1+\alpha}{1-\alpha}}\right],
\end{align}
and 
\begin{equation}\label{eq:c12closedform}
c_{12} = \frac{\lambda} {4\mu}\left[{\psi_{\alpha^2}(\tfrac{1}{2})+\log (1-\alpha^2)}\right],
\end{equation}
where $\psi_{\alpha}(x)$ is the $q$-digamma function of $x$ with $q=\alpha$. These solutions are valid for $0<\alpha< 1$. They are not defined at $\alpha=0$ or $\alpha=1$. 

In their range of validity, the closed-form expressions above can be used for calculation of the coefficients by utilising the existing special function library of numerical software such as \textit{Mathematica}~\cite{Mathematica}. Built-in convergence tests in the software can optimise the number of terms required for accuracy in calculating the $q$-digamma function. We use the closed-form solutions for calculating relative errors when deciding on the number of terms and the range of use of the classical solutions above and the asymptotic solutions discussed below.

Both classical and closed-form solutions show a slow down in convergence when the spheres are near each other, i.e. $\mu << 1$. In the near regime ($\mu \lesssim 1$) the following asymptotic solutions provide fast and accurate computation~\cite{Banerjee2019-vf}:
\begin{align}\label{eq:AsymptoticC11}
c_{11} \approx 
	\frac{\lambda}{4\mu} &\, \Bigg[\log\frac{1}{\mu}-\psi(x)-\sum_{k=1}^{K}\frac{2^{4k-1} B_{2k} (x) B_{2k}\! \left(\frac{1}{2}\right)}{(2k)! \, k}\mu^{2k} -2\pi \sin (2\pi x) \, e^{-\frac{\pi^2}{\mu}}\Bigg],\\
c_{22} \approx 
	\frac{\lambda}{4\mu}&\, \Bigg[\log\frac{1}{\mu}-\psi(y)-\sum_{k=1}^{K}\frac{2^{4k-1} B_{2k} (y) B_{2k}\! \left(\frac{1}{2}\right)}{(2k)! \, k}\mu^{2k}-2\pi \sin (2\pi y) \, e^{-\frac{\pi^2}{\mu}}\Bigg],
\end{align}
and
\begin{align}\label{eq:AsymptoticC12}
	c_{12}
	&\approx -\frac{\lambda}{4\mu } \Bigg[\log\frac{1}{\mu}+\gamma-\sum_{k=1}^{K}\frac{2^{4k-1} B_{2k} B_{2k}\! \left(\frac{1}{2}\right)}{(2k)! \,k}\mu^{2k}\Bigg],
\end{align}
where $\psi=\psi_{\alpha=1}$ is the digamma function, $\gamma=0.5772...$ is the Euler's constant, and $B_{2k}$, $B_{2k}(x)$ are the $2k$\hspace{0.01in}th Bernoulli number and Bernoulli polynomial~\cite{Abramowitz1972-nf}. 

Using the cut-off $K\simeq \pi^2/2\mu$ leads to optimally low errors in capacitances at any given $\mu$~\cite{Banerjee2019-vf}. Using $K=5$ is a practical cut-off for $\mu \le 0.35$ (or $\alpha \ge 0.5$) with errors of order $10^{-10}$ or less. Note that using too large a $K$ value makes the errors worse. For $\mu=0.35$, the optimal $K$ is $\pi^2/0.7 \simeq 14$. The $\exp{(-\pi^2/\mu)}$ term in $c_{11}, c_{22}$  matters only if the optimum $K$ is used. In this paper, with $K=5$ as cut-off, we omit that $\exp{(-\pi^2/\mu)}$ term as it is smaller than the error.

\section{Capacitance derivatives}
\label{sec:capacitance_derivatives}
The electrostatic force in Eq.~(\ref{eq:fvdefinition}) is a linear combination of the derivatives of the capacitance coefficients with respect to the sphere separation. In this section we calculate these derivatives of the capacitance coefficients discussed in Sec.~\ref{sec:capacitance}.
\subsection{Preliminary derivatives}
It is convenient to first calculate the derivatives of the main quantities used in expressing the capacitance coefficients. These derivatives allow us to express the derivatives of the capacitance coefficients in a more compact manner.

The variables $\alpha$ and $\mu$ are mathematical measures of the distance between the spheres. Their derivatives with respect to $s$ are
\begin{align}
\frac{d\alpha}{ds} =-\frac{4\alpha}{\lambda} ~~~\text{and}~~~\frac{d\mu}{ds}= \frac{2}{\lambda}.
\end{align}
The capacitance coefficients in the previous section are expressed in terms of $\mu$ and $\alpha$. The derivatives above allow us to relate the $\mu$ and $\alpha$ derivatives of the capacitance to the derivatives with respect to $s$.
The derivative of $\lambda$ with respect to $s$ is
\begin{align}
\frac{d\lambda}{ds}=4 \left(\frac{1+\alpha^2}{1-\alpha^2}\right)-\frac{\lambda}{ s}=4 \coth 2\mu-\frac{\lambda}{ s}.
\end{align}
The variation of $x,y$ with respect to $\alpha$ and $\mu$ are
\begin{align}
         x'(\alpha)&=-\frac{ (1-2x)}{4 \alpha \mu}+\frac{r\cosh [(1-2x) \, \mu ]^2}{(1+\alpha )^2 \mu },~~~~x'(\mu)=-2\alpha x'(\alpha),\\
        y'(\alpha)&=-x'(\alpha),~~~~ y'(\mu)=-x'(\mu).
\end{align}
These derivatives of $x$ and $y$ go to zero as $\alpha \rightarrow 1$ or $\mu \rightarrow 0$. 

\subsection{Classical capacitance derivatives}
Differentiating the classical result for capacitance $c_{11}$ in Eq.~(\ref{eq:c11-Classical}) with respect to $s$ gives
\begin{align}\label{eq:dc11classical}
\frac{dc_{11}}{ds}=&\frac{c_{11} }{\lambda} \frac{d\lambda}{ds}-4 \sum_{n=0}^\infty \frac{ \left(\alpha ^{3n+3x}+\alpha ^{n+x}\right) \left[n+x+\alpha  x'(\alpha ) \log \alpha  \right]}{\left(\alpha ^{2n+2x}-1\right)^2},
\end{align}
and for capacitance $c_{12}$ in Eq.~(\ref{eq:c12-Classical}) we get
\begin{align}\label{eq:dc12classical}
\frac{dc_{12}}{ds}
&=\frac{c_{12} }{\lambda} \frac{d\lambda}{ds}+4 \sum_{n=1}^\infty \frac{ \left(\alpha ^{3n}+\alpha^{n}\right) n}{\left(\alpha ^{2n}-1\right)^2}.
\end{align}
The derivative of $c_{22}$ can be obtained by replacing $x$ with $y$ and $c_{11}$ with $c_{22}$ in Eq.~(\ref{eq:dc11classical}). These derivatives of the classical results converge well when the spheres are relatively far away. Only  terms through $n=10$ are needed for a numerical accuracy of order $10^{-8}$ for $\alpha \le 0.15$ (or $\mu \ge 0.95$). The errors are calculated using the closed-form derivatives discussed below.

\subsection{Closed-form capacitance derivatives}
Differentiating the closed-form expression for $c_{11}$ in Eq.~(\ref{eq:c11closedform}) with respect to $s$ we have
\begin{align}\label{eq:dc11closedform}\nonumber
\frac{dc_{11}}{ds} &= -\frac{2\alpha} {\mu} \left[\alpha\, \psi_{\alpha^2}^{(1,0)}(x)-\psi_{\alpha}^{(1,0)}(x)+ \frac{1}{1-\alpha^2} \right]-c_{11}\left(\frac{2}{\lambda\mu}+\frac{1}{s }-\frac{4\,\coth 2 \mu}{\lambda}\right)\\
&~~~~-\frac{\alpha x'(\alpha)} {\mu}\left[ \psi_{\alpha^2}^{(0,1)}(x)-2\psi_{\alpha}^{(0,1)}(x)\right],
\end{align}
where the superscript $(0,1)$ indicates the derivative with respect to $x$ and $(1,0)$ indicates the derivative with respect to $\alpha$ or $\alpha^2$ depending on the case. 
Similarly,
\begin{align}
\frac{dc_{12}}{ds} &=-\frac{2\alpha^2}{\mu}\left[ \psi_{\alpha^2}^{(1,0)}\left(\tfrac{1}{2}\right)- \frac{1}{1-\alpha^2} \right]-c_{12}\left(\frac{2}{\lambda\mu}+\frac{1}{s }-\frac{4\,\coth 2 \mu}{\lambda}\right).
\end{align}
The derivative of $c_{22}$ can be obtained by replacing $x$ with $y$ and $c_{11}$ with $c_{22}$ in Eq.~(\ref{eq:dc11closedform}). These derivatives are defined at all points except $\alpha=0$ and $\alpha=1$.

Note that  the $\alpha$ derivative  of $\psi_\alpha(x)$ is numerically unstable in {\it Mathematica}~\cite{Mathematica} for some parts of the range $0<\alpha < 1$. To avoid inaccuracies in calculating the capacitance derivatives above, we use the inversion symmetry~\cite{Krattenthaler1996-ev}
\begin{align}\label{eq:psi_inv}
\psi_\alpha(x) &= \left(x-\tfrac{3}{2}\right) \log \alpha + \psi_{1/\alpha}(x)
\end{align} 
as an alternate definition for $\psi_\alpha(x)$ within the software. 

In principle, these closed-form expressions for the capacitance derivatives can be used to calculate the electrostatic force between the spheres for all $0<\alpha < 1$. However, derivative calculations of the $q$-digamma function encounter numerical errors near $\alpha=0$ and convergence problems near $\alpha =1$ . Therefore, for fast and accurate force calculations, we use these derivatives  only in the mid-distance range, i.e., $0.15 \leq \alpha \leq 0.5$, which corresponds to $0.95 \ge \mu \ge 0.35$.

\subsection{Asymptotic capacitance derivatives}
Differentiating the asymptotic expression for $c_{11}$ in Eq.~(\ref{eq:AsymptoticC11}) gives
\begin{align}\label{eq:dc11asymptotic} \nonumber
&\frac{dc_{11}}{ds}	
	\approx -\frac{1}{2\mu^2}+\frac{c_{11}}{\lambda}\left(\frac{{d\lambda}}{ds}-\frac{2}{\mu} \right)
		-\frac{x'(\mu)}{2\mu }\, \Bigg[ \psi'(x)+\sum_{k=1}^{K}\frac{2^{4k} B_{2k-1} (x) B_{2k}\! \left(\frac{1}{2}\right)}{(2k)!}\mu^{2k}\Bigg]\\
	-&\sum_{k=1}^{K}\frac{2^{4k-1} B_{2k} (x) B_{2k}\!  \left(\frac{1}{2}\right)}{(2k)!}\mu^{2k-2}-\frac{\pi^2 e^{-\frac{\pi^2}{\mu}}}{\mu^3} \left[ \pi \sin (2\pi x) +2\mu^2 \cos (2 \pi  x)\, x'(\mu)\right],
\end{align}
and differentiating asymptotic $c_{12}$ in Eq.~(\ref{eq:AsymptoticC12}) we get
\begin{align}\label{eq:dc12asymptotic}
   \frac{dc_{12}}{ds}	
	\approx&\frac{1}{2\mu^2}+\frac{c_{12}}{\lambda}\left(\frac{{d\lambda}}{ds}-\frac{2}{\mu} \right) +\sum_{k=1}^{K}\frac{2^{4k-1} B_{2k} B_{2k}\! \left(\frac{1}{2}\right)}{(2k)!}\mu^{2k-2}.\end{align}
The asymptotic derivative of $c_{22}$ are obtained by replacing $x$ with $y$ and $c_{11}$ with $c_{22}$ in Eq.~(\ref{eq:dc11asymptotic}).

The asymptotic derivatives are be useful for fast and accurate calculations in the near range ($\mu \lesssim 1$) provided the cut-off $K$ is carefully chosen~\cite{Banerjee2019-vf}. For practical considerations we choose $K=5$ which gives the derivatives accurately to within order $10^{-8}$ for $\mu \le 0.35$ (or $\alpha \ge 0.5$). The $\exp(-\pi^2/\mu)$ term Eq.~(\ref{eq:dc11asymptotic}) is omitted in this approximation as discussed in Sec.~\ref{sec:capacitance}.

\section{Energy and force at constant voltage} \label{sec:voltage}
In this section we analyse the electrostatic force when both spheres are held at constant voltages. The force in this case is a linear combination of the capacitance derivatives discussed in Sec.~\ref{sec:capacitance_derivatives}. We combine the classical, closed-form, and asymptotic expressions to create an accurate description of the force at all separations between the spheres. 
\subsection{Dimensionless energy and force}
\label{sec:FV}
We define a dimensionless form of the electrostatic energy as 
\begin{align}
w_V\equiv \frac{W_V}{\pi \epsilon (R_1 +R_2) V_1^2}
=& \, c_{11}+ 2 c_{12} v +c_{22} {v}^{2},
\end{align}
where $v=V_2/V_1$ is the voltage of the second sphere relative to the first sphere. We assume that $V_1$ is the larger of the two voltages so that $-1 \le v \le 1$. Since one of the two spheres has to be at a non-zero voltage for the problem to be meaningful, normalising the energy using the larger voltage ensures that there is no division by zero.

We now define a dimensionless  version of the force in  Eq.~(\ref{eq:fvdefinition}) as
\begin{align}
f_V\equiv\frac{F_V}{\pi\epsilon V_1^2}
= \frac{dc_{11}}{ds}   + 2v \frac{dc_{12}}{ds} +{v}^{2}\frac{dc_{22}}{ds} ,
\end{align}
for which the capacitance derivatives are calculated in Sec.~\ref{sec:capacitance_derivatives}. 
As defined, the dimensionless force at constant voltage satisfies the following symmetry under the interchanging of the spheres and their voltages
\begin{align}\label{eq:fvsymmetry}
\frac{f_V (r,v)}{v} = \left[\frac{f_V (r,1/v)}{1/v}\right]_{r=-r}.
\end{align}
All possible voltage and size scenarios can be covered by analysing the full range of the size asymmetry ratio $-1 <r<1$. The equations are valid for $|v|>1$ as well but don't provide any extra information as shown by the symmetry above.

For analysis presented in this paper, it is useful to explicitly calculate the force to order $\mu^2$ in the asymptotic limit so that the coefficients only depend upon the size asymmetry and the voltage ratio. Using the asymptotic derivatives in Sec. \ref{sec:capacitance_derivatives} we get
\begin{align} \nonumber
f_V &= -\frac{(1-v)^2}{2\mu^2} + \frac{(1-v)^2 (1+3r^2)}{6} \log \frac{1}{\mu}-\frac{(1-3r^2)(1+v^2)+4v}{36}\\ \nonumber
&-\frac{(1+3r^2)}{3} \phi_v(y_0)+\frac{(r-r^3)}{3} \phi_v^\prime(y_0)-(1-v)^2\frac{1+60r^2-45r^4}{90} \mu^2 \log \frac{1}{\mu}\\ \nonumber
&-\bigg[\frac{(33+330 r^2-515 r^4)(1+v^2)+8 (13+25 r^2)v}{3600}-\frac{1+60r^2-45r^4}{45}\phi_v(y_0)\\
&\hspace{0.18in}+\frac{(19r-70r^3+51r^5)}{90}\phi_v^\prime(y_0)+\frac{(r-r^3)^2}{18}\phi_v^{\prime\prime}(y_0)\bigg]\,\mu^2+{\cal O} \left(\mu^4\log \mu\right),
\label{eq:fvasymptotic}
\end{align}
where $\phi_v(x)\equiv [v^2\psi(x)+\psi(1\!-\!x)+2v\,\gamma]/2$ is a voltage weighted digamma function. 

Note that the first term in Eq.~(\ref{eq:fvasymptotic}) is always attractive and independent of the size asymmetry. It agrees with Lekner's calculation of the force at constant voltage~\cite{Lekner2012-pr}. The second term is always repulsive, symmetric under $r \rightarrow -r$, and increases with size asymmetry. The ${\cal O}(1)$ term changes under $r \rightarrow -r$ due to the presence of $\phi_v^\prime$ which involves the trigamma function.

For the numerical calculations presented in this paper (see Fig.~\ref{fig:fvr1over3} for example) we use the closed-form expressions for the middle region $0.35\le \mu \le 0.95 $ which corresponds to $0.5 \ge \alpha \ge 0.15$. In the near region $\mu\le 0.35$ (or $\alpha \ge 0.5$) we use the asymptotic capacitance and derivatives with a cut-off of $K=5$, and in the far region $\mu \ge 0.95$ (or $\alpha \le 0.15$) we use the classical capacitances and their derivatives with terms through $n=10$. 

If the $q$-digamma function is unavailable then the asymptotic solutions (with $K=5$) can be used for $\mu \le 0.6$ ($\alpha \ge 0.3$) and the classical solutions (with $n=10$) thereafter with relative errors of $10^{-5}$ or less. The regions of strong overlap in the three forms of solutions lets us accurately calculate the force between the spheres at any distance with a high degree of confidence. 

\subsection{Force $f_V$ when $R_1>R_2$}
\label{sec:FvR1gtR2}

\begin{figure}
\centering
\scalebox{0.145}{\includegraphics{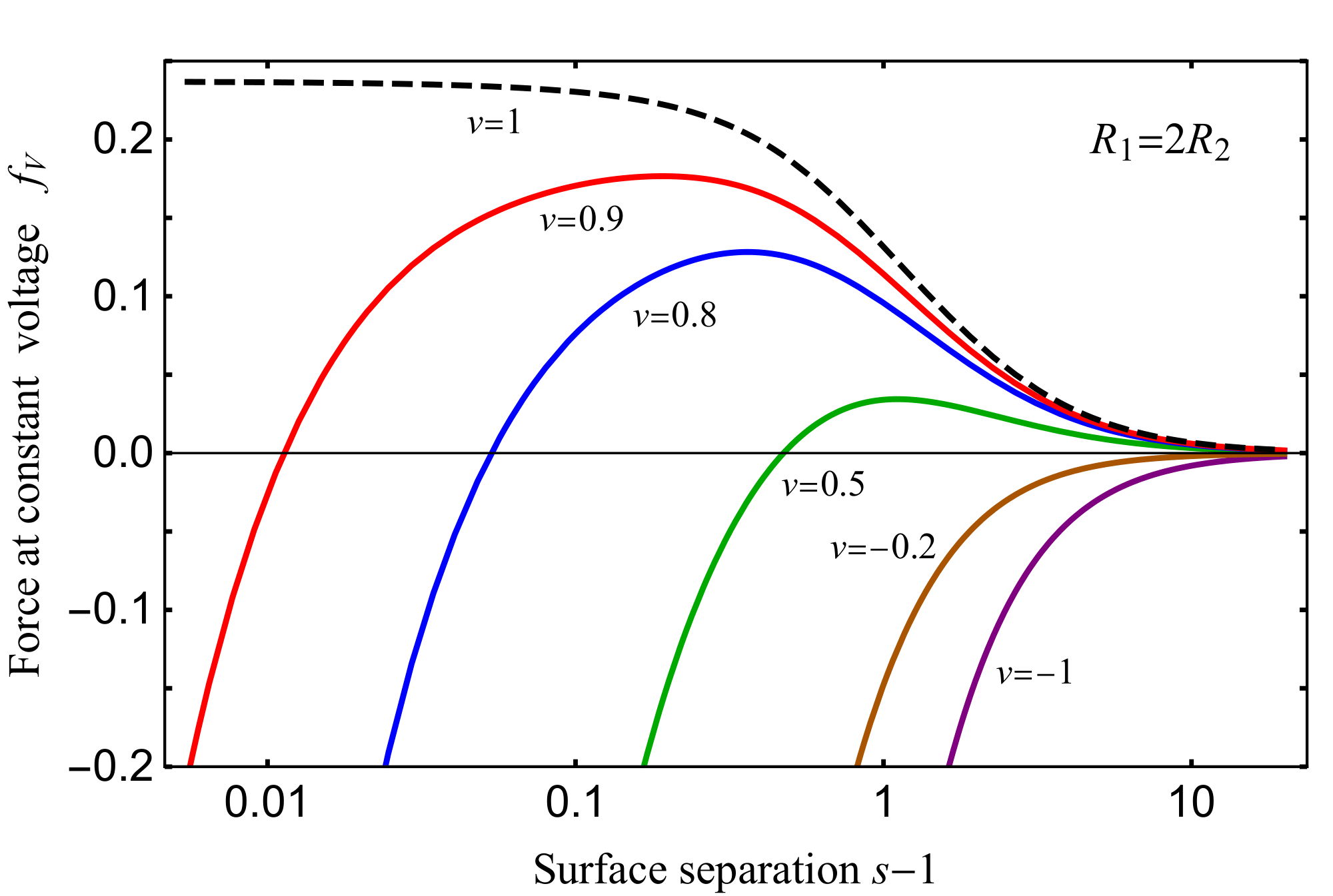}}
\caption{The dimensionless force at constant voltage, $f_V$, is plotted versus the relative surface separation, $s-1$, of the two spheres for $r=\frac{1}{3}$ (or when $R_1=2R_2$). The force is attractive (negative) at sufficiently small distances for all voltage ratios $v\ne 1$. The force is always repulsive (positive) when the two spheres have the exact same voltage, $v=1$. This repulsion at 
equal voltages decreases monotonically with increasing sphere separation.}
\label{fig:fvr1over3}
\end{figure}

We first analyse the case where the first sphere, with the higher voltage, is larger in size as well. In Fig.~\ref{fig:fvr1over3} we plot the force between the two spheres for $r=\frac{1}{3}$ which corresponds to the case where $R_1 = 2 R_2$. As expected, the force is attractive at sufficiently small sphere separation when the two spheres are not at the exact same voltage. The force is repulsive at all distances only when the two spheres have the exact same voltage, i.e., $v=1$. This attraction at sufficiently small separation when $v\ne 1$ is understood mathematically from Eq.~(\ref{eq:fvasymptotic}) - the leading term is negative and eventually dominates all the other terms. The behaviour of the force versus separation in Fig.~\ref{fig:fvr1over3} is qualitatively similar to that of equal-sized spheres held at constant voltages~\cite{Banerjee2017-xl}.

To qualitatively understand the attraction between spheres with like but unequal voltages, consider the shortest field line (along the two centres) that connects the surfaces of the two spheres. Such a field line must exist when the second sphere is at a lower voltage. At the end of this field line there must be negative charge even though the second sphere has a positive voltage. The distance between this negative charge and the positive charge at the beginning of the field line goes to zero as the two spheres are about to touch. This attractive force eventually overcomes any repulsion between the spheres. There is no such field line only when the two spheres are at the exact same voltage. In this case the force is finite even at zero separation  because the like charges move away from each other.

\begin{figure}
\centering
\scalebox{0.145}{\includegraphics{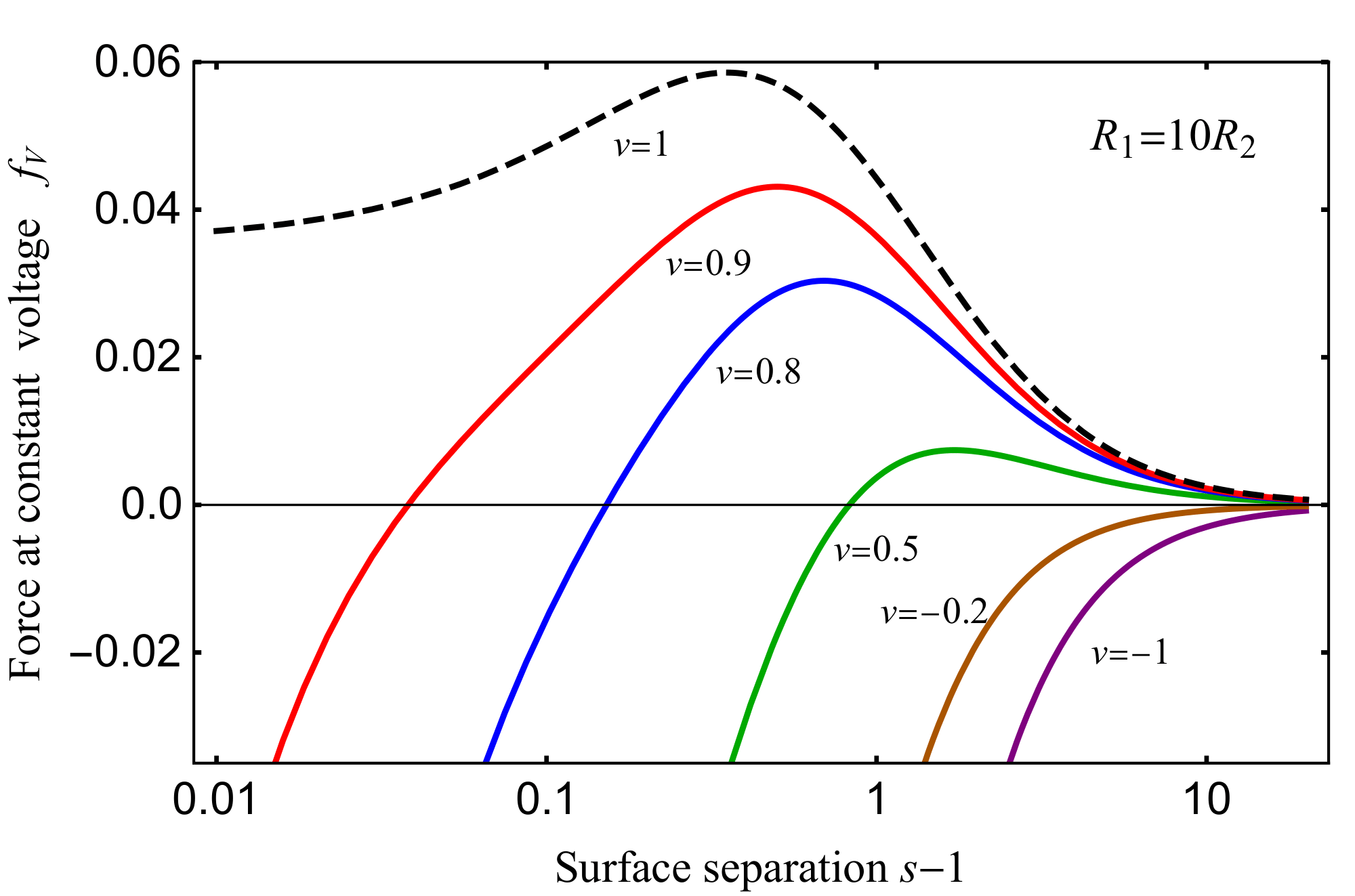}}
\caption{The dimensionless force at constant voltage, $f_V$, is plotted versus the relative sphere surface separation, $s-1$, for $r=\frac{9}{11}$  (when $R_1=10 R_2$). Similar to the $r=\frac{1}{3}$ case, the force is attractive at sufficiently small distances for all voltage ratios $v\ne 1$ and repulsive at all distances when $v=1$. However, unlike the $r=\frac{1}{3}$ case, here the repulsion at $v=1$ {\it increases} with separation for small surface separations.}
\label{fig:fvr9over11}
\end{figure}

In Fig.~\ref{fig:fvr9over11} we plot the force between the two spheres for $r=\frac{9}{11}$ which corresponds to the highly asymmetric case where $R_1 = 10 R_2$. Even in this case the force is attractive at sufficiently small sphere separation when $v \ne 1$ due to the same reasons discussed above. The force is repulsive at all distances only when $v=1$. This repulsive force, however, shows non-monotonic behaviour versus sphere separation. The repulsion initially  {\it increases} as the spheres move away from each other at contact and reaches a maximum of $1.65$ times the value at contact at $s-1=0.353$ before eventually decreasing.

We may qualitatively understand this increase in the repulsion as the spheres move away from contact by noting that the smaller sphere gains $2.45$ times its charge at contact when separation increases to $s-1=0.353$, and the charge on larger sphere decreases only to $98.9\%$ of its original value. The product of the two charges therefore increases by a factor of $2.42$. However, since the separation between the charges increases as well, the force increases only by a factor of $1.65$. Note that the first two leading order terms in Eq.~(\ref{eq:fvasymptotic}) go to zero at $v=1$. We analyse this anomalous repulsion in more detail again in Sec.~\ref{sec:normalizedrepulsion}.

\subsection{Force $f_V$ when $R_1<R_2$}
\begin{figure}
\centering
\scalebox{0.145}{\includegraphics{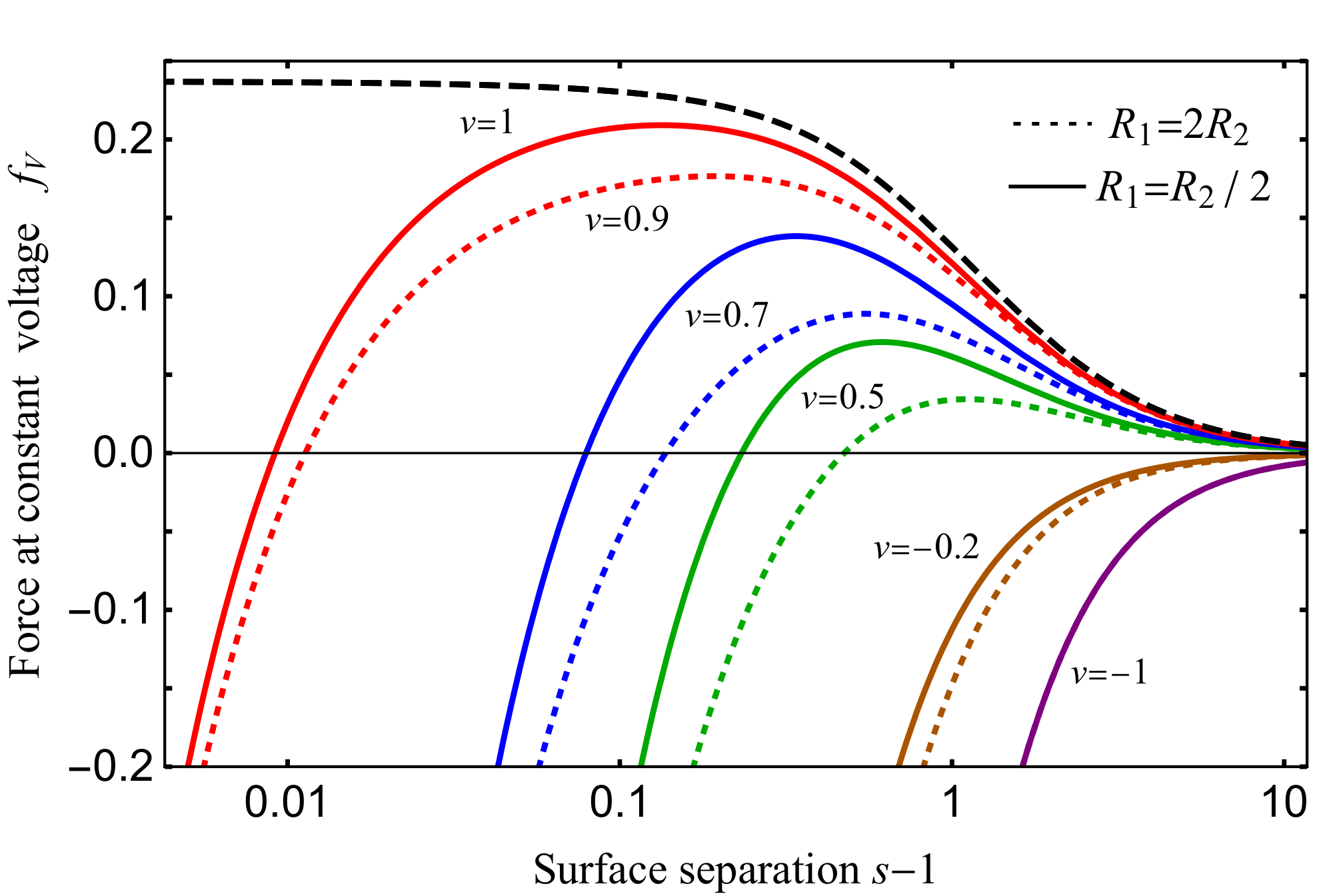}}
\caption{The dimensionless force, $f_V$, is plotted versus the relative surface separation, $s-1$, for $r=-\frac{1}{3}$ (solid lines), i.e., when $R_1 = \frac{1}{2}R_2$. The behaviour is similar to the $r=\frac{1}{3}$ case (dotted lines), except that the polarisation effect is weaker for the solid lines. This effect can be seen by comparing the switch to attraction which happens at comparatively smaller separation for the same voltage ratio.}
\label{fig:fvrnegative1over3}
\end{figure}

We now examine the case where sphere 1 with the higher voltage has a smaller radius than sphere 2. In Fig.~\ref{fig:fvrnegative1over3} we plot the force between the two spheres for $r=-\frac{1}{3}$ which corresponds to the case when $R_1 = \frac{1}{2}R_2$. The force is attractive at sufficiently small sphere separation when the two spheres are not at the exact same voltage. The force is repulsive at all distances only when the two spheres have the exact same voltage, i.e., $v=1$. The polarisation effect, which causes the attraction at small separation, is weaker in this case than for $r=\frac{1}{3}$. This makes sense if we note that the smaller sphere at lower voltage loses charge rapidly with decreasing distance to maintain its given voltage in the presence of the  larger sphere.  In comparison, the charge on the lower voltage larger sphere does not change much in the vicinity of the smaller sphere at a higher voltage.

\begin{figure}
\centering
\scalebox{0.145}{\includegraphics{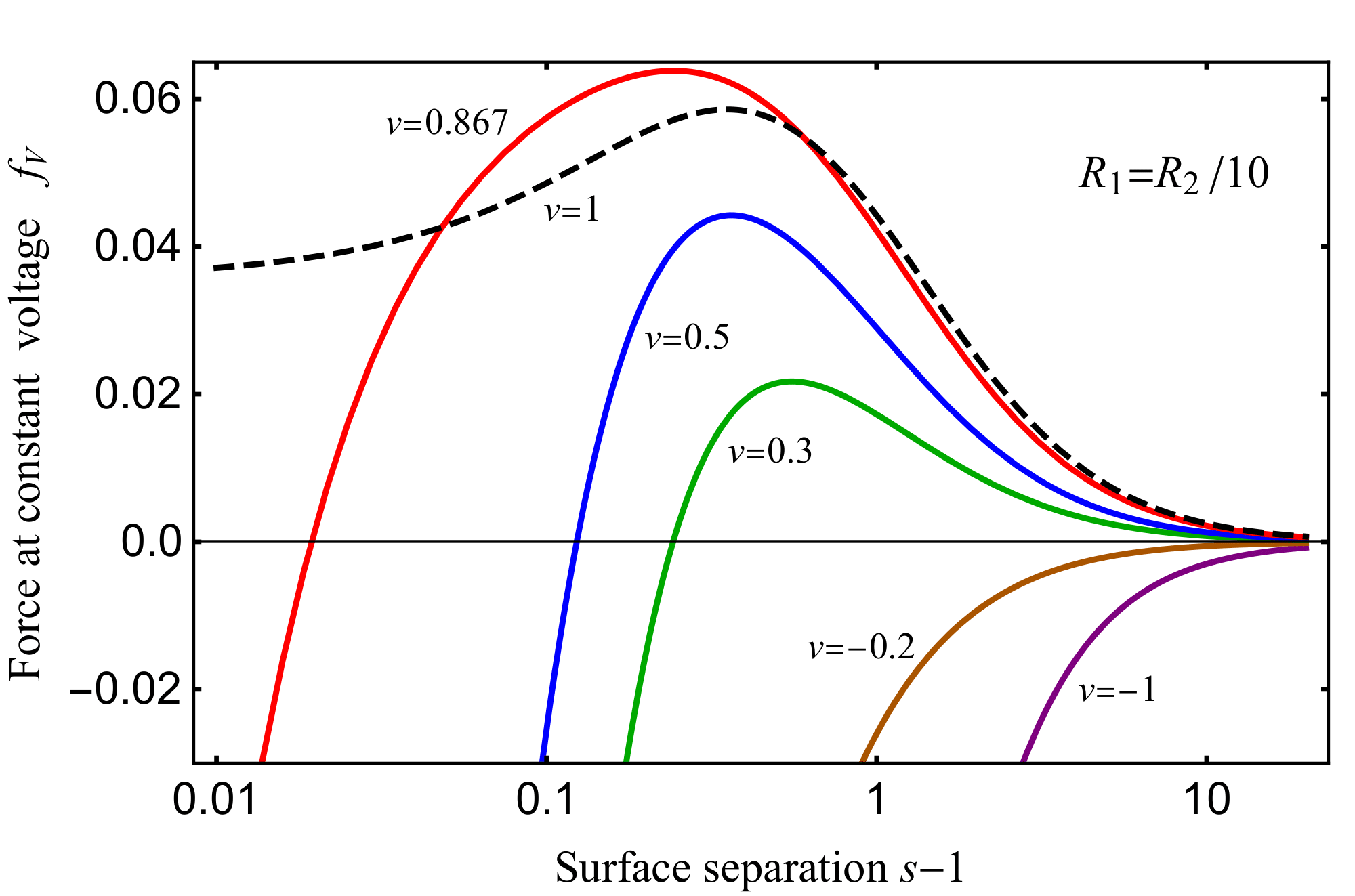}}
\caption{The dimensionless force, $f_V$,  is plotted  versus the relative surface separation, $s-1$, for $r=-\frac{9}{11}$ (i.e., $R_1 = \frac{1}{10}R_2$). The force plots are similar to the $r=\frac{9}{11}$ case, with one notable exception: for some separations, decreasing the voltage on the larger sphere can {\it increase} the repulsion. The ``optimal" $v=0.867...$ maximises the relative vertical ``jump" from the $v=1$ curve over all voltage ratios and sphere separations (see Table \ref{table:forceinversion} for details).}
\label{fig:fvrnegative9over11}
\end{figure}

In Fig.~\ref{fig:fvrnegative9over11} we plot the force between the spheres for $r=-\frac{9}{11}$ which corresponds to case where $R_1 =\frac{1}{10} R_2$. The force shows similar behaviour to the $r=\frac{9}{11}$ case in Fig.~\ref{fig:fvr9over11}. The $v=1$ curves are identical by definition since when the two spheres have equal voltage, either one can be chosen as sphere 1. However, unlike the  $r=\frac{9}{11}$ case, at certain distances, it is possible to increase the repulsive force between the two spheres by {\it decreasing } the voltage of the larger sphere! The value $v=0.867...$ is calculated numerically to maximise the relative vertical jump from the $v=1$ curve over all voltages and separations.

How can sphere with less voltage exert a greater repulsion? Calculations show that decreasing the voltage to $86.7\%$ on the larger sphere at $s-1=0.112$ reduces its charge by $15\%$ but the smaller sphere acquires $61\%$ more charge to maintain its voltage. The product of the two charges thus increases by $37\%$. However, we hypothesise that the extra charge on the smaller sphere pushes the charge on the larger sphere further away and the effective distance between the two charge distributions increases. Additionally, there is now some negative charge on the larger sphere as well which causes some attraction. Thus, there is an overall increase of $18\%$  in the repulsive force.

In Table~\ref{table:forceinversion} we list such force anomalies for several different sizes of the two spheres. For every $r$ value we numerically calculate the optimum lowering of the voltage from $v=1$  and the optimum separation, $s-1$, that causes the maximum relative increase in the force. Note that when the two spheres are not at equal voltage, there are some field lines from sphere 1 that end up on sphere 2 and cause some attraction between charges of opposite polarity. Even then, the lower voltage case has overall more repulsion than the equal voltage case.

\renewcommand{\arraystretch}{1.4}
\begin{table} 
\centering
\begin{tabular}{||c|c|c| c|c|c|c||}
\hline
$\hspace{0.1in}r \hspace{0.1in}$ & \!\!$R_1:R_2$\!\! & \!optimal $v$\! &  $s\!-\!1$  & \!\!$f_V(v \!=\! v_{opt})$\!\! &  \!$f_V(v \!=\! 1)$\!  & \!\!$f_V(v_{opt})/f_V(1)$\!\!  \\ [0.5ex]

 \hline\hline

--\,$\nicefrac{1}{2}$ & $1:3$ & 0.978 & 0.0790 & 0.175 & 0.174 & 1.0056\\
\hdashline[4pt / 1pt]
--\,$\nicefrac{3}{5}$ & $1:4$ & 0.951& 0.103 & 0.138 & 0.135 & 1.022\\
\hline
--\,$\nicefrac{2}{3}$ & $1:5$ & 0.929 & 0.112 & 0.113 & 0.108 & 1.045\\
\hdashline[4pt / 1pt]
--\,$\nicefrac{5}{7}$ & $1:6$ & 0.912 & 0.116 & 0.0961 & 0.0896 & 1.072\\
\hline
--\,$\nicefrac{3}{4}$ & $1:7$ & 0.897 & 0.116 & 0.0831 & 0.0756 & 1.099\\
\hdashline[4pt / 1pt]
--\,$\nicefrac{7}{9}$ & $1:8$ & 0.885 & 0.115 & 0.0731 & 0.0649 & 1.128\\
\hline
--\,$\nicefrac{4}{5}$ & $1:9$ & 0.875 & 0.114 & 0.0653 & 0.0565 & 1.156\\
\hdashline[4pt / 1pt]
--\,$\nicefrac{9}{11}$ & $1:10$ & 0.867 & 0.112 & 0.0589 & 0.0497 & 1.184\\
[0.5ex]
\hline

\end{tabular}

\caption{The data shows that the repulsive force is {\it greater} when the larger sphere is at an optimal {\it lower} voltage $f_V(v\!=\!v_{opt})$ than when both spheres are at the same voltage, $f_V(v\!=\!1)$. Also listed is the optimal separation at which both forces are calculated. See Fig. \ref{fig:fvrnegative9over11} for example. The voltage and separation are optimised for achieving the largest ratio (last column) of the two forces. }
\label{table:forceinversion}
\end{table}

To calculate the minimum size asymmetry where lowering the voltage can increase the force, let the voltage ratio be $v = 1 - \varepsilon$, where $\varepsilon<<1$. Substituting this voltage ratio in the expression for force we get
\begin{equation}
\begin{aligned}
f_V&=\frac{d c_{11}}{ds} +2(1 - \varepsilon)\frac{d c_{12}}{ds}+ (1-\varepsilon)^2 \frac{d  c_{22}}{ds} \\
&=f_V(v\!=\!1)-2\varepsilon \left( \frac{d c_{12}}{ds} +\frac{d  c_{22}}{ds}\right)+{\cal O}(\varepsilon^2).
\end{aligned}
\end{equation}
So the criterion for the force to increase when the voltage is reduced is that the coefficient of $\varepsilon$ has to be positive at some separation.
The conditions for the critical case are given by
\begin{align}
\frac{d }{ds}(c_{12}+c_{22})=0~~~\text{and}~~~\frac{d^2 }{ds^2}(c_{12}+c_{22})=0.
\end{align}
Solving for the two conditions numerically gives
$r = -0.3226...$ and $s-1 \rightarrow 0$. Thus, the critical case is when two spheres with $R_1 \simeq \frac{1}{2} R_2$ are about to touch each other. Note that the data in Table~\ref{table:forceinversion} alludes to this critical point as well.

Analysing the capacitance coefficients near $\mu=0$ yields 
\begin{align}
\frac{c_{12}+c_{22}}{1-r^2} =& -\frac{1}{2} \left[\gamma + \psi \left(y_0\right)\right] +\frac{\mu^2}{24}\times \\ \nonumber
&\left[-1+r^2- (2+6r^2)[\gamma + \psi(y_0)]+2r(1-r^2)\psi^{\prime} (y_0)\right]+{\cal{O} }(\mu^4),
\end{align}
where $y_0 = (1+r)/2$ is the value of $y$ as $\mu\rightarrow 0$. Plotting the coefficient of $\mu^2$ shows that it is negative for $r < -0.3226...$ which corresponds to the size ratio $R_1 \simeq \frac{1}{2} R_2$. That is, the repulsion can be higher at lower voltage if the first sphere is smaller than about half the size of the second sphere. Note that $\mu^2 \sim (s-1)$ in this limit.

\section{Energy and force at constant charge}\label{sec:charge}
In this section we analyse the electrostatic force when the charges on the spheres are held constant. In addition to the standard definition discussed in Sec.~\ref{sec:introduction}, we develop another alternative formulation for the force which allows us to use our results from the constant voltage case. We combine the classical, closed-form, and asymptotic expressions to create an accurate description of the force at all separations between the spheres.

\subsection{Dimensionless formulation}

We define a dimensionless form of the electrostatic energy  as
\begin{equation}
w_Q   \equiv \frac{4\pi\epsilon(R_1+R_2) W_Q}{q_0 Q_1^2}= \frac{1}{q_0} \left(p_{11}  + 2 p_{12} q + p_{22} q^2\right), 
\end{equation}
where $p_{ij}\equiv 2\pi\epsilon (R_1+R_2)P_{ij}$ are dimensionless, $q=Q_2/Q_1$ is the charge ratio between the spheres, and
\begin{align}
q_{0}
&=\frac{\gamma+\psi(y_0 )}{\gamma+\psi(x_0)}=\frac{\phi_{1}(y_0)-\frac{\pi}{2}\cot\pi y_0}{\phi_{1}(y_0)+\frac{\pi}{2}\cot\pi y_0}
\end{align}
is the charge ratio at contact with $\phi_{1}=\phi_{v=1}$. In the ratio $q_0$ above, $x_0 = (1-r)/2$ is the value of $x$ as $\mu\rightarrow 0$. Note that $y_0=1-x_0$. 

This normalisation of $W_Q$ ensures that the dimensionless energy and force at contact are the same regardless of which sphere is designated as $1$. Sphere $1$ is chosen such that $q_0 \, |Q_1|\ge |Q_2|$ to ensure that $|V_1|\ge |V_2|$ near contact. Additionally, sphere $1$ is chosen to be of positive polarity without loss of generality since the energy and the force remain the same if we switch polarities of both spheres. 

We can now calculate the dimensionless force at constant charge as
\begin{align}\label{eq:fq} \nonumber
f_Q&\equiv\frac{4\pi\epsilon(R_1+R_2)^2}{q_0 Q_1^2} F_Q =
  -\frac{4\pi\epsilon(R_1+R_2)^2}{q_0 Q_1^2}\frac{dW_Q}{dS}\\
  &=-\frac{dw_Q}{ds}= -\frac{1}{q_0}\left(\frac{dp_{11}}{ds}   + 2 q \frac{dp_{12}}{ds}  +q^2 \frac{dp_{22}}{ds}\right). 
\end{align}
As defined, the dimensionless force at constant charge satisfies the symmetry
\begin{align}
f_Q (r,q_0) = \left[f_Q (r,q_0)\right]_{r=-r}
\end{align}
at $q=q_0$ and more generally for any $q$,
\begin{align}\label{eq:fqsymmetry}
 \frac{q_0 f_Q (r,q)}{q} = \left[  \frac{q_0 f_Q (r,1/q)}{1/q}\right]_{r=-r}.
\end{align}
All possible charge and size scenarios are covered by analysing the parameter space $-q_0\le q \le q_0$ along with $-1<r<1$. Although we restrict ourselves to $|q|\le q_0$, the force expressions  hold for $|q|>q_0$ as well. However, those cases do not provide any additional information than that already discussed in the paper. Any $|q|>q_0$ is equivalent to a case with ratio $q\rightarrow 1/q$ (and thus $|q|<q_0$) and $r \rightarrow -r$ in our parameter space 
as shown by the symmetry above.

\subsection{Alternate formulation of force at fixed charge}
The derivatives of the coefficients of potential $p_{ij}$ are nontrivial to calculate and simplify. Here, we rewrite $f_Q$ in terms of the force at constant voltage discussed earlier in Sec.~\ref{sec:FV} in the manner below.

Imagine disconnecting the batteries at a certain distance $s$. At this point the spheres now are at fixed charge. But the force before and after disconnecting the batteries should not change at the same $s$ value since the charge on the sphere stays the same. Setting $F_V=F_Q$ and replacing $V_1$ in terms of its corresponding charge values gives
\begin{align}
f_V=\frac{F_V}{\pi\epsilon V_1^2}
= \frac{F_Q}{\pi\epsilon (P_{11} Q_1 + P_{12}Q_2)^2}
=\frac{4\pi \epsilon (R_1+R_2)^2 F_Q}{ Q_1^2 (p_{11} + p_{12}q)^2}.
\end{align}
Therefore
\begin{equation}\label{eq:fq-of-fv}
\begin{aligned}
\frac{4\pi \epsilon (R_1+R_2)^2 F_Q}{q_0 Q_1^2 }=
f_Q=& \frac{(p_{11} + p_{12}q )^2}{q_0} f_V.
\end{aligned}
\end{equation}
Note that $f_V$ is in terms of $v$ which should be replaced by its equivalent in terms of $q$,
\begin{equation}
v = \frac{V_2}{V_1} = \frac{ p_{22}q + p_{12}}{p_{11} + p_{12}q} = \frac{c_{11}\,q-c_{12} }{c_{22} - c_{12}\, q},
\end{equation}
since the charges are now fixed whereas the voltage ratio $v$ varies with distance.  After expressing all the $p_{ij}$ in terms of $c_{ij}$ and comparing the coefficients of the derivatives of $c_{ij}$, it is easily verified that the $f_Q$ expressions in Eqs.~(\ref{eq:fq})~and~(\ref{eq:fq-of-fv}) are equal to each other.

To understand the plots presented below it is useful to expand this force at constant charge in the limit $\mu \rightarrow 0$ as 
\begin{align}\label{eq:fqmu}
f_Q &= - \frac{2\left(1-\frac{q}{q_0}\right)^2 (1-\xi^2)}{(1-r^2)^2 \mu^2 \left[2\log\frac{1}{\mu}-\phi_{-1}(y_0)+\xi^2 \phi_1(y_0)\right]^2}\\ \nonumber
&+\frac{\left[\left(1\!+\!\frac{q}{q_0}\right)+\xi  \left(1\!-\!\frac{q}{q_0}\right)\right]^2 \left[(1\!-\!r^2) (2 r \phi_1'(y_0)\!-\!1)-(2\!+\!6 r^2) \phi_1(y_0)\right]}{6 \left(1\!-\xi ^2\right) \left(1-r^2\right)^2 \phi_1(y_0)^2}
+ {\cal O} \left(\!\frac{1}{\log {\mu}}\!\right), 
\end{align}
with $\xi= \pi \cot(\pi y_0)/2\phi_1(y_0)$ where $\phi_{\pm 1} = \phi_v$ at $v=\pm 1$. Note that the leading term in Eq.~(\ref{eq:fqmu}) is always negative.
Thus for $q\ne q_0$ this leading term always dominates the higher order terms and causes an overall attractive (negative) force for sufficiently small $\mu$, which agrees with Lekner's result~\cite{Lekner2012-pr}. The second term, dominant when $q=q_0$, is always repulsive and approaches a finite limit as $\mu \rightarrow 0$. All higher order terms go to zero in this limit.

\subsection{Calculations of force $f_Q$} 
We now analyse the electrostatic force between the two spheres at constant charge using the expressions developed above. The sphere designated as 1 is the more ``positive" of the two and does not develop any negative charge density even upon polarisation. The second sphere develops some negative charge density before contact (because of its lower voltage) even if it has an overall positive charge. To cover all possible charge and size scenarios, we examine both $r>0$ ($R_1>R_2$) and $r<0$ $(R_1 < R_2)$ cases since even the smaller sphere can be the more positive of the two.

\begin{figure}
\centering
\scalebox{0.145}{\includegraphics{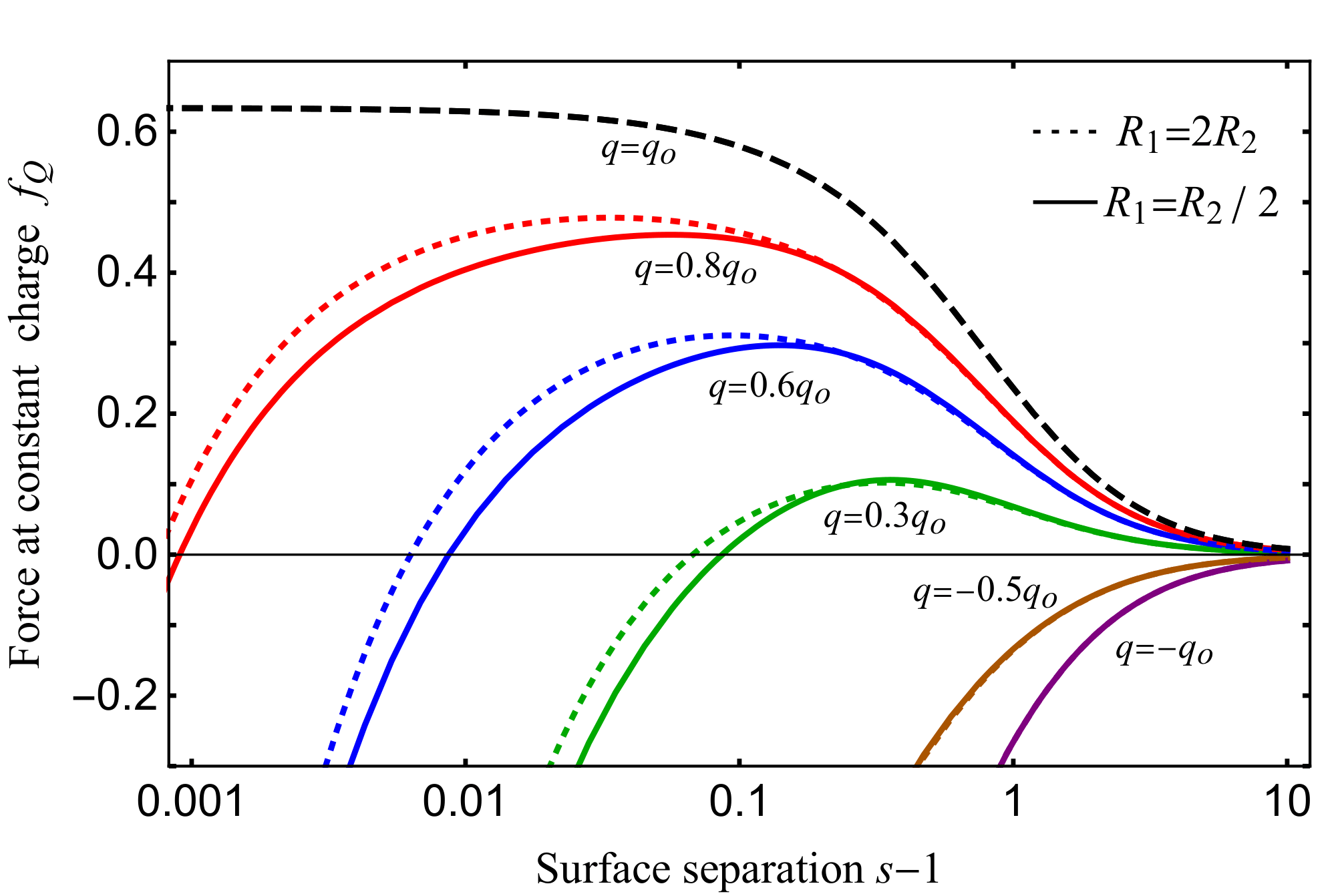}}
\caption{The dimensionless force, $f_Q$, is plotted versus the relative surface separation, $s-1$, for $r=-\frac{1}{3}$ (solid lines), i.e., $R_1=\frac{1}{2}R_2$. Also plotted is the $r=\frac{1}{3}$ case (dotted lines) for comparison. The force is attractive at sufficiently small distances for all charge ratios when $q < q_0$ and always repulsive only when the two spheres have the charge ratio at contact, $q=q_0$. The polarisation effect that causes attraction at small separation is stronger when the smaller sphere is the more ``positive" of the two.}
\label{fig:fqr1over3}
\end{figure} 

In Fig.~\ref{fig:fqr1over3} we plot the force when one sphere is twice the size of the other, i.e., $r=\pm \frac{1}{3}$. In both cases the spheres attract each other at sufficiently small separations unless their charge ratio is exactly the same as the charge ratio at contact, i.e., $q = q_0$. This attraction due to charge polarisation is stronger for $r=-\frac{1}{3}$ (solid lines), where the smaller sphere is the more positive of the two.
The behaviour of the force versus separation in Fig.~\ref{fig:fqr1over3} is qualitatively similar to that of equal-sized spheres with fixed charges~\cite{Banerjee2017-xl}.

The ``perfect" charge ratio, $q_0$, ensures that both spheres are at the same voltage right before contact and there are no electric field lines between them. Thus, there is no negative charge density on either sphere which can cause any attraction. When one of the spheres has less charge than this ideal ratio, there is at least one field line from the sphere with the greater (than perfect) charge to the sphere with less (than perfect) charge along the line joining the two. The opposite ends of this field line have opposite charge density. When regions of opposite charge density get sufficiently close before contact, their attraction overcomes the repulsion between the like charges to cause an overall attraction between the spheres.

\begin{figure}
\centering
\scalebox{0.145}{\includegraphics{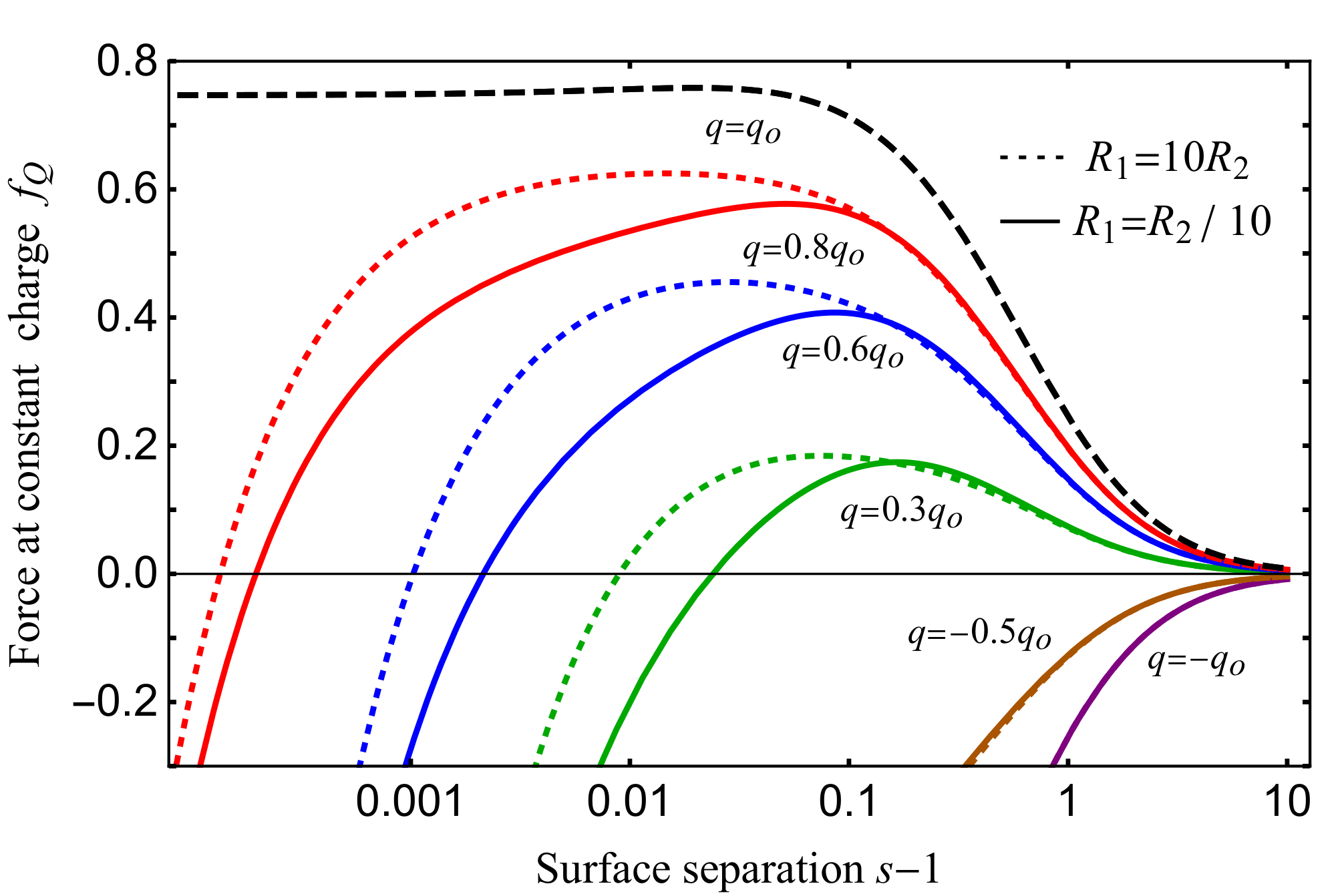}}
\caption{The dimensionless force, $f_Q$, is plotted versus the relative surface separation, $s-1$, for $r=\frac{9}{11}$ (dotted lines) and $r=-\frac{9}{11}$ (solid lines). The force is attractive at sufficiently small distances for all charge ratios $q < q_0$ and repulsive at all distances only when the two spheres have $q=q_0$, the charge ratio at contact. This plot is similar to that in Fig. \ref{fig:fqr1over3} with one exception: the repulsion at $q=q_0$ {\it increases} as the two spheres separate away from contact before eventually decreasing.}
\label{fig:fqr9over11}
\end{figure} 

In Fig.~\ref{fig:fqr9over11} we plot the force when one sphere is ten times the other, i.e., $r=\pm \frac{9}{11}$. In both cases the spheres attract each other at sufficiently small separations unless their charge ratio is exactly the same as the charge ratio at contact, $q = q_0$. The features of this plot are similar to those for the $r=\pm \frac{1}{3}$ case in Fig.~\ref{fig:fqr1over3} with one exception: similar to $v=1$ in the constant voltage case, the repulsive force when $q = q_0$ is not maximum when the two spheres are in contact and {\it increases} when the two spheres separate.

To understand how two spheres can repel more at a finite separation than when at contact, we hypothesise that the charge on the larger sphere moves back towards the point of contact when the two spheres separate. This charge rearrangement must be the main mechanism that causes a stronger horizontal push along the axis of symmetry since neither sphere gains any charge. 

Comparing the force plots for constant charge with the force plots at constant voltage in Sec.~\ref{sec:voltage}, we see that polarisation effects in constant charge and constant voltage are in the opposite directions; that is, the dashed curves lie on opposite sides of the solid curves where the force switches to attraction. This contrasting behaviour is due to the difference in main mechanism of charge polarisation in the two cases. 

In the constant voltage case the main mechanism that drives the force characteristics is the gaining/losing of charge by the smaller sphere. This mechanism is stronger when the larger sphere has more voltage. In comparison, the main mechanism of polarisation in the constant charge case is the redistribution of charge on the larger sphere. This charge redistribution is stronger when the smaller sphere has a greater than its fair share of the charge and is able to cause more charge separation on the larger sphere. 

Another contrast between the constant charge and the constant voltage cases is the value of the separation at which the repulsion switches to attraction. This separation is much smaller in the charge case for similar numerical value of the voltage and charge ratios. By expanding the variable voltage ratio, $v$, for a fixed charge ratio $q$ we see that
\begin{align}
v= 1-\frac{q \psi\left(\frac{1-r}{2}\right)-\gamma  (1-q)-\psi\left(\frac{1+r}{2}\right)}{(1+q) \log \frac{1}{\mu }+\gamma  q-\psi\left(\frac{1+r}{2}\right)} +{\cal O}\left(\frac{\mu^2}{\log{\mu}}\right).
\end{align}
Note that the voltage ratio $v$ goes to $1$ as the separation $\mu \rightarrow 0$ regardless of the charge ratio $q$. Hence, much smaller separations are needed to overcome the $v=1$ like repulsion.

\section{Repulsion increase with separation}
\label{sec:normalizedrepulsion}
The anomalous behaviour of the electrostatic force at contact, where the repulsion increases as the spheres separate, happens only for sufficiently large size asymmetries. Similar behaviour is observed in both constant voltage and constant charge cases although the force increase  is much larger in the constant voltage case. In this section we analyse this repulsion increase with separation between the two spheres as a function of their size asymmetry.

For the constant voltage case, setting $v=1$ and expanding $f_V$ for small $\mu$ gives
\begin{align}\nonumber
 f_V (v\!=\!1)=& -\left(\frac{1}{3}+r^2\right)\phi_1(y_0)+\left(\frac{1-r^2}{6}\right)\Big[2 r\,\phi_1^\prime(y_0) - 1\Big]+\frac{\mu^2}{360}\times\\ \label{eq:fvmu2}
 & \bigg[-\left(17+86 r^2-103 r^4\right)+\left(8+480 r^2-360 r^4\right) \phi_1(y_0) \\ \nonumber
&\hspace{0.0in}- \left(76 r-280 r^3+204 r^5\right) \phi_1^\prime(y_0)-20 \left(r-r^3\right)^2 \phi_1^{\prime\prime}(y_0)\bigg]+{\cal O}\left(\mu^4\right).
\end{align}
Setting the coefficient of $\mu^2$ to zero and numerically solving for the critical asymmetry ratio yields $r_c=0.423...$ ($R_1\simeq 2.5 R_2$). For $|r|>r_c$, the coefficient of $\mu^2$ is positive and the force increases with separation starting at contact! As discussed earlier in Sec.~\ref{sec:FvR1gtR2}, this increase in force is mainly due to the increase in charge on the smaller sphere.

\begin{figure}
\centering
\scalebox{0.145}{\includegraphics{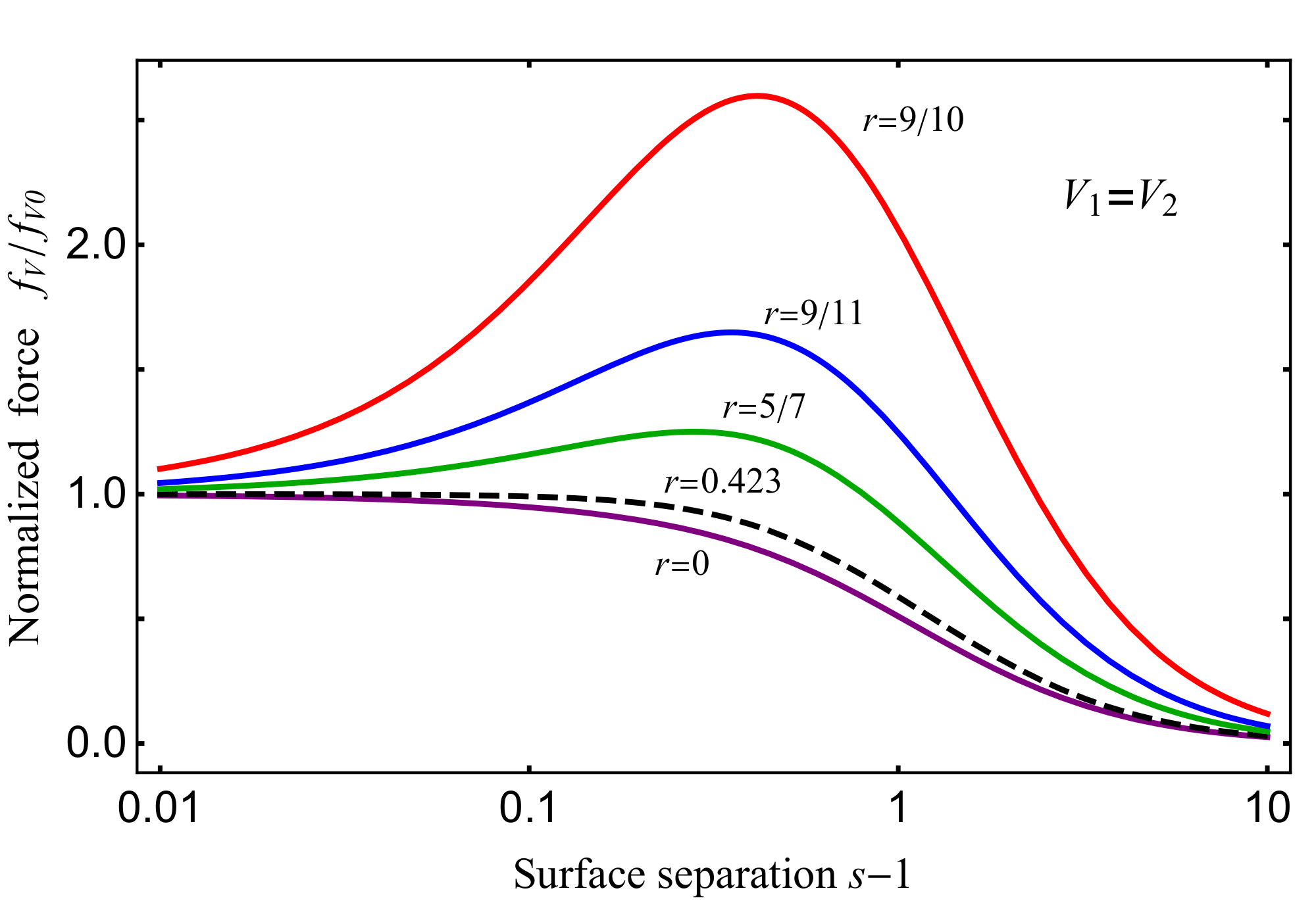}}
\caption{The force at constant voltage, $f_V$, normalised by the force at contact, $f_{V0}$, is plotted against the surface separation, $s-1$, for the voltage ratio $v=1$. Above a critical size asymmetry, $r_c=0.423...$ ($R_1 \simeq 2.5 R_2$), the force {\it increases} as the two spheres separate from contact. For $r=0.9$ the force increases to 2.6 times its value at contact at $s-1=0.41$ (see Table. \ref{table:repulsionatcontact}).}
\label{fig:NormalizedFV}
\end{figure} 

In Fig.~\ref{fig:NormalizedFV} we highlight this non-monotonic behaviour of the repulsive force with sphere separation as a function of the size asymmetry. To compare all the cases we normalise the force by its value at contact. For size ratios greater than about 5:2, the repulsion increases at first as the spheres move away from each other from contact. We observe that the peak of the normalised force increases monotonically with $r$.

By using $f_V(v\!=\!1)$ in Eq.~(\ref{eq:fvmu2}) and setting $q=q_0$ the force at constant charge for small $\mu$ yields 
\begin{align}\label{eq:fqmu2}
f_Q (q\!=\!q_0) &= \frac{4f_V(v\!=\!1)}{(1-r^2)^2[\phi_1(y_0)^2-(\pi^2/4)\cot^2(\pi y_0)]}\times\\\nonumber
&\left [1-\mu^2\frac{1-r^2+(2+6r^2)\,\phi_1(y_0)-2(r-r^3)\,\phi_1^\prime(y_0)}{6\,\phi_1(y_0)}  \right]+{\cal O}\left(\frac{\mu^2}{\log \mu}\right).
\end{align}
Isolating the coefficient of $\mu^2$ above and setting it to zero gives $r_c=0.4872...$(i.e., $R_1 \simeq 3 R_2$). The force at contact, $f_{Q0}$, in its dimensionless form is the same as the Kelvin factor calculated in Ref.~\cite{Lekner2012-ua}.

\begin{figure}
\centering
\scalebox{0.145}{\includegraphics{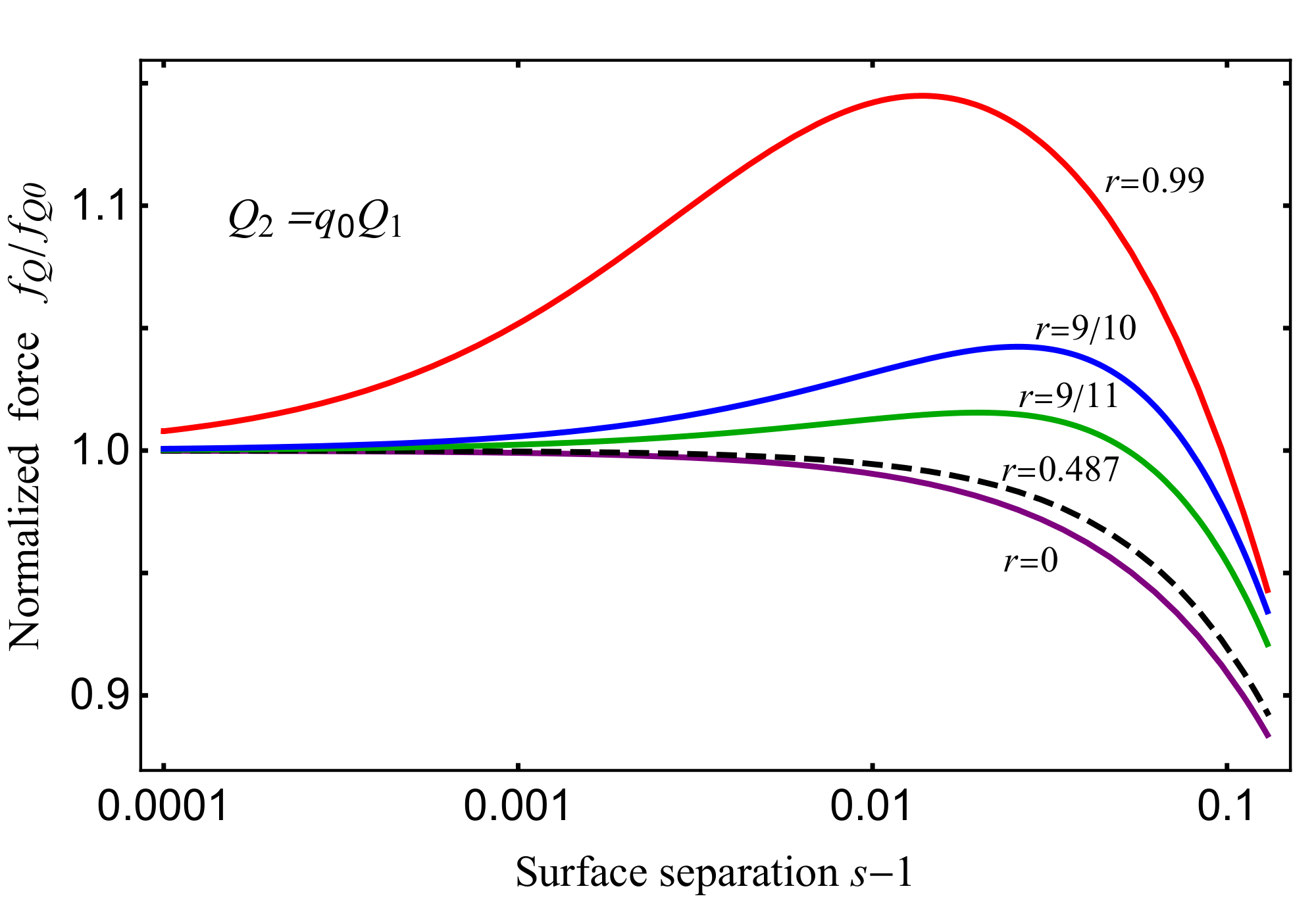}}
\caption{The force $f_Q$ normalised by the force at contact, $f_{Q0}$, is plotted against separation $s-1$ for the charge ratio $q_0$. Above a critical size asymmetry, $r_c=0.487...$ ($R_1 \simeq 3 R_2$), the force {\it increases} as the two spheres move away from contact. For $r=0.9$ the force increases to 1.042 times its contact value (see Table.~\ref{table:repulsionatcontact}). The increase in repulsion here is much smaller than that in Fig.~\ref{fig:NormalizedFV}. High size asymmetries (see $r=0.99$) are needed for a significant ($14\%$) force increase.}
\label{fig:NormalizedFQ}
\end{figure}

In Fig.~\ref{fig:NormalizedFQ} we plot $f_Q$ normalised by its value at contact for several different values of $r$. Compared to the constant voltage case in Fig.~\ref{fig:NormalizedFV}, the normalised force shows a much smaller increase and the peak happens at much smaller separations. Since neither sphere gets any additional charge, the only mechanism by which the force can increase is the redistribution of the charge which creates a greater 
horizontal component of the force as separation increases.

In Table~\ref{table:repulsionatcontact} we list values of the maximum normalised repulsion between the spheres and the corresponding separations at which this maximum happens. This data can be tested against computer simulations or experimental measurements. The constant voltage case appears to be a more likely candidate for experimental verification since the normalised force values are much larger than in the constant charge case. Also, it is easier to maintain two spheres at constant voltage than constant charge due to the variety of mechanisms through which the charge can dissipate from objects that are not perfectly isolated.

\renewcommand{\arraystretch}{1.4}
\begin{table}
\centering
\begin{tabular}{||c|c|c| c|c|c||}
\hline
$\hspace{0.1in}r \hspace{0.1in}$ &$R_1:R_2$ &\!\! Max \!$f_V/f_{V0}$ & $s-1$ \!\!&\! \! Max \!$f_Q/f_{Q0}$ \!\!&  $s-1$ \\ [0.5ex]

 \hline\hline

$\nicefrac{1}{2}$ & $3:1$ & $1.014$ & 0.100 &--&--\\
\hdashline[4pt / 1pt]
$\nicefrac{3}{5}$ & $4:1$ & $1.077$ & 0.194 & --&--\\
\hline
$\nicefrac{2}{3}$ & $5:1$ & $1.159$ & 0.244 & -- &--\\
\hdashline[4pt / 1pt]
$\nicefrac{5}{7}$ & $6:1$ & $1.250$ & 0.279 & 1.003 &0.00890\\
\hline
$\nicefrac{3}{4}$ & $7:1$ & $1.346$ & 0.304 & 1.006 &0.0127\\
\hdashline[4pt / 1pt]
$\nicefrac{7}{9}$ & $8:1$ & $1.445$ & 0.324 & 1.009& 0.0157\\
\hline
$\nicefrac{4}{5}$ & $9:1$ & $1.546$ & 0.340 & 1.012&0.0180\\
\hdashline[4pt / 1pt]
$\nicefrac{9}{11}$ & $10:1$ & $1.648$ & 0.353 & 1.015 & 0.0198\\
\hline
$\nicefrac{6}{7}$ & $13:1$ & $1.961$ & 0.382 & 1.025 & 0.0233\\
\hdashline[4pt / 1pt]
$\nicefrac{15}{17}$ & $16:1$ & $2.278$ & 0.401 & 1.034 & 0.0249\\
\hline
$\nicefrac{9}{10}$ & $19:1$ & $2.597$ & 0.415 & 1.042 & 0.0257\\
[0.5ex]
\hline
\end{tabular}
\caption{The data shows that the repulsion between the two spheres {\it increases} as they move away from contact. The value of the force is listed as a ratio to the value of the force  at contact, $f_V/f_{V0}$ for constant voltage, and $f_Q/f_{Q0}$ for constant charge. Also listed is the separation $s-1$ at which this force ratio is a maximum. The values $f_{V0}$ and $f_{Q0}$ are given by the $\mu \rightarrow 0$ limits of Eqs.~(\ref{eq:fvmu2})~and~(\ref{eq:fqmu2}).}
\label{table:repulsionatcontact}
\end{table}

\section{Summary}
The main new results presented in this paper are the asymptotic expansions and closed-form expressions for the electrostatic force between two conducting spheres.  We supplement these results with the well known classical method-of-images series in the far region for best convergence and computational speed at all distances while maintaining high accuracy.

These force expressions allow us to investigate the force as a function of the sphere size asymmetry. Our calculations reveal some new results when the size asymmetry between the spheres is large. We highlight these results by proposing the following three experiments:

{\it Experiment 1.} Let one sphere be 19 times larger than the other (see $r=9/10$ curve in Fig.~\ref{fig:NormalizedFV}). Hold both spheres at equal voltages and measure the electrostatic force at contact. Now move the spheres to 1.42 times this distance of closest approach while keeping their voltages the same. The measured force should now be 2.60 times larger.

{\it Experiment 2.} Let one sphere be 10 times larger than the other (see Fig.~\ref{fig:fvrnegative9over11}). Hold both spheres at equal voltages and measure the electrostatic force at a centre-to-centre distance of $1.112(R_1+R_2)$. Now decrease the voltage of the larger sphere to $86.7\%$ of the original value while keeping the separation fixed. The measured force should now be $18.4\%$ larger.

{\it Experiment 3.} Let one sphere be 19 times larger than the other (see $r=9/10$ curve in Fig.~\ref{fig:NormalizedFQ}). Charge both spheres and measure their electrostatic force at contact. Now move the spheres to 1.026 times this distance of closest approach while keeping their charges fixed. The measured force should now be 1.042 times larger. This force increase is similar to but much smaller than that in experiment 1.

The main mechanism for the force increase in the first two experiments is the substantial gaining of charge by the smaller sphere. This mechanism is missing in the third experiment since the charges on both spheres are fixed. We hypothesise that the mechanism behind the force increase in the third experiment is the rearrangement of charges on the larger sphere which decreases the effective distance between the spheres even as their centre-to-centre distance increases.

The three experiments listed above highlight the anomalous behaviour of the electrostatic force for high sphere size asymmetries. Additional details of such behaviour are provided in Tables~\ref{table:forceinversion}~and~\ref{table:repulsionatcontact}. For convenience we work with dimensionless form of the electrostatic force. If needed, the dimensionless forms can be converted back to SI units through the relations
\begin{align}
F_V = \pi \epsilon V_1^2 f_V~~~\text{and}~~~ F_Q = \frac{q_0 Q_1^2}{4\pi\epsilon(R_1+R_2)^2} f_Q.
\end{align}
In the paper we designate sphere 1 to be stronger than sphere 2 in order to put a bound on the voltage and charge ratios and avoid repetition of cases. However, there is no such requirement; all relations hold as long as $V_1$ and $Q_1$ are nonzero.

\section*{Acknowledgments}

We thank the Mac Armour Fellowship for support.  
\bibliography{main.bib}

\begin{thebibliography}{22}
\providecommand{\natexlab}[1]{#1}
\providecommand{\url}[1]{\texttt{#1}}
\expandafter\ifx\csname urlstyle\endcsname\relax
  \providecommand{\doi}[1]{doi: #1}\else
  \providecommand{\doi}{doi: \begingroup \urlstyle{rm}\Url}\fi

\bibitem[Thomson(1872)]{Thomson1872-jh}
William Thomson.
\newblock On the mutual attraction or repulsion between two electrified
  spherical conductors.
\newblock \emph{Reprint of Papers on Electrostatics and Magnetism}, pages
  86--97, 1872.
\newblock \doi{10.1017/cbo9780511997259.007}.

\bibitem[Maxwell(1954)]{Maxwell1954-ka}
James~Clerk Maxwell.
\newblock {IX} and {XI}.
\newblock In \emph{A Treatise on Electricity and Magnetism}, volume~1. Dover
  Publishing, 3 edition, 1954.

\bibitem[Russell(1927)]{Russell1927-lr}
A~Russell.
\newblock The electrostatic problem of two conducting spheres.
\newblock \emph{J. Inst. Electr. Eng.}, 65:\penalty0 517--535(18), May 1927.
\newblock ISSN 0099-2887.
\newblock \doi{10.1049/jiee-1.1927.0053}.

\bibitem[Lekner(2011)]{Lekner2011-pz}
John Lekner.
\newblock Capacitance coefficients of two spheres.
\newblock \emph{J. Electrostat.}, 69:\penalty0 11--14, February 2011.
\newblock \doi{10.1016/j.elstat.2010.10.002}.

\bibitem[Lekner(2012{\natexlab{a}})]{Lekner2012-pr}
John Lekner.
\newblock Electrostatic force between two conducting spheres at constant
  potential difference.
\newblock \emph{J. Appl. Phys.}, 111\penalty0 (7):\penalty0 076102, April
  2012{\natexlab{a}}.
\newblock ISSN 0021-8979.
\newblock \doi{10.1063/1.3702438}.

\bibitem[Lekner(2012{\natexlab{b}})]{Lekner2012-ua}
John Lekner.
\newblock Electrostatics of two charged conducting spheres.
\newblock \emph{Proc. R. Soc. A}, 468\penalty0 (2145):\penalty0 2829--2848,
  2012{\natexlab{b}}.
\newblock \doi{10.1098/rspa.2012.0133}.

\bibitem[Kolikov et~al.(2012)Kolikov, Ivanov, Krastev, Epitropov, and
  Bozhkov]{Kolikov2012-zq}
Kiril Kolikov, Dragia Ivanov, Georgi Krastev, Yordan Epitropov, and Stefan
  Bozhkov.
\newblock Electrostatic interaction between two conducting spheres.
\newblock \emph{J. Electrostat.}, 70\penalty0 (1):\penalty0 91--96, February
  2012.
\newblock ISSN 0304-3886.
\newblock \doi{10.1016/j.elstat.2011.10.008}.

\bibitem[Meyer(2015)]{Meyer2015-fd}
Matthias Meyer.
\newblock Numerical and analytical verifications of the electrostatic
  attraction between two like-charged conducting spheres.
\newblock \emph{J. Electrostat.}, 77:\penalty0 153--156, October 2015.
\newblock ISSN 0304-3886.
\newblock \doi{10.1016/j.elstat.2015.08.007}.

\bibitem[Banerjee et~al.(2017)Banerjee, Levy, Davis, and
  Wilkerson]{Banerjee2017-xl}
Shubho Banerjee, Mason Levy, Mckenna Davis, and Blake Wilkerson.
\newblock Exact and approximate capacitance and force expressions for the
  electrostatic interaction between two equal-sized charged conducting spheres.
\newblock \emph{IEEE Trans. Ind. Appl.}, PP:\penalty0 1--1, February 2017.
\newblock ISSN 0093-9994.
\newblock \doi{10.1109/TIA.2017.2672744}.

\bibitem[Lekner(2016)]{Lekner2016-df}
John Lekner.
\newblock Regions of attraction between like-charged conducting spheres.
\newblock \emph{Am. J. Phys.}, 84\penalty0 (6):\penalty0 474--477, 2016.
\newblock ISSN 0002-9505.
\newblock \doi{10.1119/1.4942449}.

\bibitem[Banerjee et~al.(2019)Banerjee, Peters, Song, and
  Wilkerson]{Banerjee2019-vf}
Shubho Banerjee, Thomas Peters, Yi~Song, and Blake Wilkerson.
\newblock Closed-form and asymptotic capacitance coefficients for the
  electrostatics of two spheres.
\newblock \emph{J. Electrostat.}, 101:\penalty0 103369, September 2019.
\newblock ISSN 0304-3886.
\newblock \doi{10.1016/j.elstat.2019.103369}.

\bibitem[Harrison et~al.(2020)Harrison, Nicoll, Ambaum, Marlton, Aplin, and
  Lockwood]{Harrison2020-xo}
R~Harrison, Keri~A Nicoll, Maarten Ambaum, Graeme Marlton, Karen~L Aplin, and
  Michael Lockwood.
\newblock Precipitation modification by ionization.
\newblock \emph{Phys. Rev. Lett.}, 124, May 2020.
\newblock ISSN 0031-9007.
\newblock \doi{10.1103/PhysRevLett.124.198701}.

\bibitem[Feng and Hays(2003)]{Feng2003-wu}
James~Q Feng and Dan~A Hays.
\newblock Relative importance of electrostatic forces on powder particles.
\newblock \emph{Powder Technol.}, 135-136:\penalty0 65--75, October 2003.
\newblock ISSN 0032-5910.
\newblock \doi{10.1016/j.powtec.2003.08.005}.

\bibitem[Cordero et~al.(2017)Cordero, Meyer, Nandwana, and
  Dehoff]{Cordero2017-dk}
Zachary~C Cordero, Harry~M Meyer, Peeyush Nandwana, and Ryan~R Dehoff.
\newblock Powder bed charging during electron-beam additive manufacturing.
\newblock \emph{Acta Mater.}, 124:\penalty0 437--445, February 2017.
\newblock ISSN 1359-6454.
\newblock \doi{10.1016/j.actamat.2016.11.012}.

\bibitem[Varadwaj et~al.(2017)Varadwaj, Varadwaj, and
  Yamashita]{Varadwaj2017-rb}
Arpita Varadwaj, Pradeep Varadwaj, and Koichi Yamashita.
\newblock Do surfaces of positive electrostatic potential on different halogen
  derivatives in molecules attract? like attracting like!
\newblock \emph{J. Comput. Chem.}, 39, December 2017.
\newblock ISSN 0192-8651.
\newblock \doi{10.1002/jcc.25125}.

\bibitem[Hudlet et~al.(1998)Hudlet, Saint~Jean, Guthmann, and
  Berger]{Hudlet1998-fe}
S~Hudlet, M~Saint~Jean, C~Guthmann, and J~Berger.
\newblock Evaluation of the capacitive force between an atomic force microscopy
  tip and a metallic surface.
\newblock \emph{Eur. Phys. J. B}, 2\penalty0 (1):\penalty0 5--10, March 1998.
\newblock ISSN 1434-6036.
\newblock \doi{10.1007/s100510050219}.

\bibitem[Law and Rieutord(2002)]{Law2002-go}
Bruce Law and Francois Rieutord.
\newblock Electrostatic forces in atomic force microscopy.
\newblock \emph{Phys. Rev. B Condens. Matter}, 66, June 2002.
\newblock ISSN 0163-1829.
\newblock \doi{10.1103/PhysRevB.66.035402}.

\bibitem[Krattenthaler and Srivastava(1996)]{Krattenthaler1996-ev}
C~Krattenthaler and H~M Srivastava.
\newblock Summations for basic hypergeometric series involving a $q$-analogue
  of the digamma function.
\newblock \emph{Comput. Math. Appl.}, 32\penalty0 (3):\penalty0 73--91, 1996.
\newblock ISSN 0898-1221.
\newblock \doi{10.1016/0898-1221(96)00114-9}.

\bibitem[Smythe(1968)]{Smythe1968-rd}
William~R Smythe.
\newblock 5.08.
\newblock In \emph{Static and Dynamic Electricity}. McGraw-Hill, 3 edition,
  1968.

\bibitem[Banerjee and Wilkerson(2017)]{Banerjee2017-wk}
Shubho Banerjee and Blake Wilkerson.
\newblock Asymptotic expansions of {L}ambert series and related $q$-series.
\newblock \emph{Int. J. Number Theory}, 13\penalty0 (08):\penalty0 2097--2113,
  2017.
\newblock ISSN 1793-0421.
\newblock \doi{10.1142/S1793042117501135}.

\bibitem[Inc.()]{Mathematica}
Wolfram~Research{,} Inc.
\newblock Mathematica, {V}ersion 12.0.
\newblock Champaign, IL, 2019.

\bibitem[Abramowitz and Stegun(1972)]{Abramowitz1972-nf}
Milton Abramowitz and Irene~A Stegun.
\newblock Chapter 23.
\newblock In \emph{Handbook of Mathematical Functions with Formulas, Graphs,
  and Mathematical Tables}. Dover, 1972.

\end{thebibliography}
\bibliographystyle{unsrtnat}

\end{document}